\documentclass[10pt]{article}
\usepackage{graphicx}
\begin{document}

\twocolumn[\begin{center}
{\large {\bf Transverse Wave Propagation in Relativistic Two-fluid Plasmas around Schwarzschild-anti-de Sitter
Black Hole}}\\
\vspace{.5cm}
M. Atiqur Rahman\\
{\it Department of Applied Mathematics, Rajshahi University, Rajshahi - 6205, Bangladesh}
\end{center}
\centerline{\bf Abstract}
\baselineskip=18pt
\bigskip
\begin{center}
\parbox{16cm}{The $3+1$ formalism of Thorne and Macdonald has been used to derive the linear two-fluid equations for transverse waves in the plasma closed to the Schwarzschild-anti-de Sitter
 (SAdS) black hole. We reformulate the relativistic two-fluid equations to take account of gravitational effects due to the event horizon and negative cosmological constant and describe the set of simultaneous linear equations for the perturbations. Using a local approximation we investigate the one-dimensional radial propagation of Alfv\'en and high frequency electromagnetic waves. We derive the dispersion relation for these waves and solve it for the wave number $k$ numerically.}
\end{center}
\vspace{0.2cm}
]\section{ Introduction}\label{sec1}
Black holes belong to the most fascinating objects predicted by Einstein's theory of gravitation. Black holes are still mysterious \cite{one}. In recent years the investigation of plasmas in the black hole environment is important because a successful study of the waves and emissions from plasmas falling into a black hole will be of great value in aiding the observational identification of black hole candidates. The theory of general relativity and its application to the plasma close to the black hole horizon, however, have remained esoteric, and little concrete astrophysical impact has been felt. Within $3R_s$ (3 Schwarzschild radii) it is possible to have plasma \cite{two,three,four,five}. The plasma in the black hole environment may act as a fluid and Black holes greatly affect the surrounding plasma medium (which is highly magnetized) with their enormous gravitational fields. Hence plasma physics in the vicinity of a black hole has become a subject of great interest in astrophysics.

Sakai and Kawata have developed the linearized treatment of plasma waves using special relativistic formulation. Such an investigation of wave propagation in general relativistic two-fluid plasma near a black hole is important for an understanding of plasma processes.
In order to investigate the behavior of plasma, a more intuitive mathematical description of general relativistic two-fluid plasma has been introduced (Buzzi, Hines, and Treumann (BHT)) \cite{six,seven} based on the 3+1 membrane paradigm (Thorne, Price, and Macdonald (TPM)) \cite{eight,nine,ten,eleven} near the horizon of the Schwarzschild black hole.
 In this paper we shall apply this method to investigate the nature of the two-fluid plasma (electron and positron or ion) near the SAdS black hole atmosphere.

The membrane paradigm is mathematically equivalent to the standard, full general relativistic theory of black holes in which the black holes event horizon are replaced with a membrane endowed (electric charge, electrical conductivity, and finite temperature and entropy) so far as physics outside the event horizon is concerned. But the formulation of all physics in this region turns out to be very much simpler than it would be using the standard covariant approach of general relativity.
In the TPM formulation, work connected with black holes has been facilitated by replacing the hole's event horizon with a membrane endowed with electric charge, electrical conductivity, and finite temperature and entropy.

The study of plasma wave in the presence of strong gravitational fields using the $3+1$ approach of is still in its early stages. Zhang \cite{twelve,thirteen} has considered the care of ideal magneto hydrodynamics waves near a Kerr black hole, accreting for the effects of the holes angular momentum but ignoring the effects due to the black hole horizon. Holcomb and Tajima \cite{fourteen}, Holcomb \cite{fifteen}, and Dettmann et al. \cite{sixteen} have considered some properties of wave propagation in a Friedmann universe.

Actually the $3+1$ approach was originally developed by Arnowitt, Deser, and Misner \cite{seventeen} to study the quantization of the gravitational field. Since then, their formulation has most been applied in studying numerical relativity \cite{eighteen}. TPM extended the $3+1$ formalism to include electromagnetism and applied it to study electromagnetic effects near the Kerr black hole.

The principal objective of this work is to solve the linearized two-fluid equations for the perturbations in the Rindler coordinate and then to determine the Alfv\'en and high frequency electromagnetic waves by using the local approximation with appropriate boundary conditions in the vicinity of Schwarzschild-anti-de Sitter Black Hole.

The solutions of black holes in Anti-de Sitter spaces come from the Einstein equations with a negative cosmological constant. Anti-de Sitter black holes are different from de Sitter black holes. The difference consisting in them is due to minimum temperatures that occur when their sizes are of the order of the characteristic radius of the anti-de Sitter space.  For larger Anti-de Sitter black holes, their red-shifted temperatures measured at infinity are greater. This implies that such black holes can be in stable equilibrium with thermal  radiation at a certain temperature. Recently the AdS/CFT correspondence suggested that String theory in anti-de Sitter space (AdS) equal to Conformal Field Theory (CFT) in one less dimension
 \cite{nineteen,twenty,twenty one} and there is a duality between quantum gravity in an asymptotically de Sitter spacetime and a CFT living on its boundary \cite{twenty two,twenty three,twenty four,twenty five,twenty six,twenty seven}. So our study on the Schwarzschild-anti-de Sitter black hole is reasonable and meaningful.

This paper is organized as follows. In section \ref{sec2} we summarize the $3+1$ formulation of general relativity. We describe the nonlinear two-fluid equations expressing continuity and conservation of energy and momentum in section \ref{sec3}. The two-fluids are coupled together through Maxwell's equations for the electromagnetic fields. For zero gravitational fields these equations reduce to the corresponding special relativistic expressions. In section \ref{sec4} we restrict one-dimensional wave propagation in the radial $z$ direction.  We linearized the two-fluid equations for transverse wave propagation with Rindler approximation in section \ref{sec5}. In section \ref{sec6}, we express the value of the unperturbed quantities. We discuss the local or mean-field approximation used to obtain numerical solutions for the wave dispersion relations. We describe the dispersion relation for the transverse waves in section \ref{sec7}, and give the numerical procedure for determining the roots of the dispersion relation. The Alfv\'en and high frequency modes are shown in sections \ref{sec8} and \ref{sec9}. Finally, in section \ref{sec10} we present our remarks. We use units $G=c=k_B=1$.

\section{Formulation of $3+1$ Spacetime}\label{sec2}

The $3+1$ formulation of general relativity developed by TPM \cite{eight,nine,ten,eleven} is based on the concept of selecting a preferred set of spacelike hyper surfaces which form the level surfaces of a congruence of timelike curves. The metric for a static spherically symmetric spacetime
with mass $M$ and a negative cosmological constant $\Lambda =-3/\ell^2 $ in asymptotically
anti-de Sitter spacetime has the form
\begin{equation}
ds^2=-f(r)dt^2+\frac{1}{f(r)}dr^2+r^2(d\theta ^2+{\rm sin}^2\theta d\varphi ^2),\label{eq1}
\end{equation}
where the metric function $f(r)$ is
\begin{equation}
f(r)=1-\frac{2M}{r}+\frac{r^2}{\ell^2},\label{eq2}
\end{equation}
and the coordinates are defined such that $-\infty\leq t\leq \infty $, $r\geq 0$, $0\leq \theta \leq \pi $ and $0\leq \phi \leq 2\pi$.
The function $f(r)$ vanished at the zeros of the cubic equation
\begin{equation}
r^3+\ell^2r-2M\ell^2=0.\label{eq3}
\end{equation}
The only real roots of this equation is
\begin{eqnarray}
r_+=\frac{2}{3}\sqrt{3}\ell\,\sinh[\frac{1}{3}\sinh^{-1}(3\sqrt{3}\frac{M}{\ell})].\label{eq4}
\end{eqnarray}
Expanding $r_+$ in terms of $M$ with $1/\ell^2<<M^2/9$, we obtain
\begin{equation}
r_+=2M(1-\frac{4M^2}{\ell^2}+ ....).\label{eq5}
\end{equation}
Therefore, we can write $r_+=2M\eta, $ with $\eta <1$. The event horizon of the SAdS black hole is smaller than the Schwarzschild event horizon, $r_H=2M$.
An absolute three-dimensional space defined by the hyper surfaces of constant universal time $t$ is described by the metric
\begin{equation}
ds^2=\frac{1}{f(r)}dr^2+r^2(d\theta ^2+{\rm sin}^2\theta d\varphi ^2)\label{eq6}.
\end{equation}
The fiducial observers (FIDO) at rest in this absolute space measured all local physical quantities using FIDO's proper time. For this FIDO use a local Cartesian coordinate system with unit basis vectors tangent to the coordinate lines
\begin{equation}
{\bf e}_{\hat r}=\sqrt{f(r)}\frac{\partial }{\partial r},\hspace{.5cm}{\bf e}_{\hat \theta }=\frac{1}{r}\frac{\partial }{\partial \theta },\hspace{.5cm}{\bf e}_{\hat \varphi }=\frac{1}{r{\rm sin}\theta }\frac{\partial }{\partial \varphi }\label{eq7}.
\end{equation}
For a spacetime viewpoint rather than a $3+1$ split of spacetime, the set of orthonormal vectors also includes the basis vector for the time coordinate given by
\begin{equation}
{\bf e}_{\hat 0}=\frac{d}{d\tau }=\frac{1}{\alpha }\frac{\partial }{\partial t}\label{eq8},
\end{equation}
where $\alpha $ is the lapse function (or redshift factor) defined by
\begin{equation}
\alpha (r)\equiv \frac{d\tau }{dt}=({1-\frac{2M}{r}+\frac{r^2}{\ell^2}})^{1/2}\label{eq9}.
\end{equation}

The gravitational acceleration felt by a FIDO is given by \cite{eight,nine,ten,eleven}
\begin{equation}
{\bf a}=-\nabla {\rm ln}\alpha =-\frac{1}{\alpha }({\frac{M}{r^2}}+{\frac{2r}{\ell^2}}){\bf e}_{\hat r}\label{eq10},
\end{equation}
while the rate of change of any scalar physical quantity or any three-dimensional vector or tensor, as measured by a FIDO, is defined by the derivative
\begin{equation}
\frac{D}{D\tau }\equiv \left(\frac{1}{\alpha }\frac{\partial }{\partial t}+{\bf v}\cdot \nabla \right)\label{eq11},
\end{equation}
$\bf v$ being the velocity of a fluid as measured locally by a FIDO.

\section{Two-fluid Equations}\label{sec3}
In this section we consider the plasma mixture of two perfect fluids either electron-positron or electron-ion and describe the two-fluid equations for continuity, the conservation of energy and momentum, and Maxwell's equation in $3+1$ formalism. Here we choose these type of fluid component because the dispersion relation that result from the following investigation are valid for either one of this two fluids since no assumption can be made upon the mass, number density, pressure or temperature of this two fluids. In the TPM $3+1$ formulation, the continuity equation for each of the fluid species is
\begin{equation}
\frac{\partial }{\partial t}(\gamma _sn_s)+\nabla \cdot (\alpha \gamma _sn_s{\bf v}_s)=0\label{eq12},
\end{equation}
where $s$ is $1$ for electrons and $2$ for positrons (or ions). For a perfect relativistic fluid of species $s$ in three-dimensions, the energy density $\epsilon _s$, the momentum density ${\bf S}_s$, and stress-energy tensor $W_s^{jk}$ are given by
\begin{eqnarray}
&&\epsilon _s=\gamma _s^2(\varepsilon _s+P_s{\bf v}_s^2),\quad{\bf S}_s=\gamma _s^2(\varepsilon _s+P_s){\bf v}_s, \nonumber\\
&&W_s^{ij}=\gamma _s^2(\varepsilon _s+P_s)v_s^jv_s^k+P_sg^{jk},\label{eq13}
\end{eqnarray}
where ${\bf v}_s$ is the fluid velocity, $n_s$ is the number density, $P_s$ is the pressure, and $\varepsilon _s$ is the total energy density defined by
\begin{equation}
\varepsilon _s=m_sn_s+P_s/(\gamma _g-1)\label{eq14}.
\end{equation}
The gas constant $\gamma _g$ take the value  $4/3$ for $T\rightarrow \infty $ and $5/3$ for $T\rightarrow 0$.
The ion temperature profile is closely adiabatic and it approaches $10^{12}\,K$ near the horizon \cite{twenty eight}. Far from the (event) horizon electron (positron) temperatures are essentially equal to the ion temperatures, but closer to the horizon the electrons are progressively cooled to about $10^8-10^9\,K$ by mechanisms like multiple Compton scattering and synchrotron radiation. Using the conservation of entropy the equation of state can be expressed as
\begin{equation}
\frac{D}{D\tau }\left(\frac{P_s}{n_s^{\gamma _g}}\right)=0.\label{eq15}
\end{equation}
where $D/D\tau =(1/\alpha )\partial /\partial t+{\bf v}_s\cdot \nabla $. The full equation of state for a relativistic fluid, as measured in the fluid's rest frame, is as follows \cite{twenty nine,thirty}:
\begin{equation}
\varepsilon =m_sn_s+m_sn_s\left[\frac{P_s}{m_sn_s}-\frac{{\rm i}H_2^{(1)^\prime }({\rm i}m_sn_s/P_s)}{{\rm i}H_2^{(1)}({\rm i}m_sn_s/P_s)}\right]\label{eq16},
\end{equation}
where the $H_2^{(1)}(x)$ are Hankel functions.
The quantities of Eq. (\ref{eq13}) for the electromagnetic field take the following form:
\begin{eqnarray}
&&\epsilon _s=\frac{1}{8\pi }({\bf E}^2+{\bf B}^2),\quad{\bf S}_s=\frac{1}{4\pi }{\bf E}\times {\bf B},\nonumber\\
&&W_s^{jk}=\frac{1}{8\pi }({\bf E}^2+{\bf B}^2)g^{jk}-\frac{1}{4\pi }(E^jE^k+B^jB^k).\quad\label{eq17}
\end{eqnarray}
The conservation of energy and momentum equations are written, respectively, as follows \cite{eight,nine,ten,eleven}:
\begin{eqnarray}
\frac{1}{\alpha }\frac{\partial }{\partial t}\epsilon _s&=&-\nabla \cdot {\bf S}_s+2{\bf a}\cdot {\bf S}_s,\label{eq18}\\
\frac{1}{\alpha }\frac{\partial }{\partial t}{\bf S}_s&=&\epsilon _s{\bf a}-\frac{1}{\alpha }\nabla \cdot (\alpha {\stackrel{\longleftrightarrow }{\bf W}}_s)\label{eq19}.
\end{eqnarray}
When the two-fluid plasma couples to the electromagnetic fields, Maxwell's equations take the following form:
\begin{eqnarray}
\nabla \cdot {\bf B}&=&0,\label{eq20}\\
\nabla \cdot {\bf E}&=&4\pi \sigma ,\label{eq21}\\
\frac{\partial {\bf B}}{\partial t}&=&-\nabla \times (\alpha {\bf E}),\label{eq22}\\
\frac{\partial {\bf E}}{\partial t}&=&\nabla \times (\alpha {\bf B})-4\pi \alpha {\bf J},\label{eq23}
\end{eqnarray}
where the charge and current densities are defined by
\begin{equation}
\sigma =\sum_s\gamma _sq_sn_s,\hspace{1.2cm}{\bf J}=\sum_s\gamma _sq_sn_s{\bf v}_s\label{eq24}.
\end{equation}
Using Eq. (\ref{eq13}), the energy and momentum conservation Eqs. (\ref{eq18}) and (\ref{eq19}) can be rewritten for each fluid species $s$ in the following form
\begin{eqnarray}
&&\frac{1}{\alpha }\frac{\partial }{\partial t}P_s-\frac{1}{\alpha }\frac{\partial }{\partial t}[\gamma _s^2(\varepsilon _s+P_s)]-\nabla \cdot [\gamma _s^2(\varepsilon _s+P_s){\bf v}_s]\nonumber\\
&&+\gamma _sq_sn_s{\bf E}\cdot {\bf v}_s+2\gamma _s^2(\varepsilon _s+P_s){\bf a}\cdot {\bf v}_s=0,\label{eq25}
\end{eqnarray}
\begin{eqnarray}
&&\gamma _s^2(\varepsilon _s+P_s)\left(\frac{1}{\alpha }\frac{\partial }{\partial t}+{\bf v}_s\cdot \nabla \right){\bf v}_s+\nabla P_s\nonumber\\
&&-\gamma _sq_sn_s({\bf E}+{\bf v}_s\times {\bf B})+{\bf v}_s\left(\gamma _sq_sn_s{\bf E}\cdot {\bf v}_s+\frac{1}{\alpha }\frac{\partial }{\partial t}P_s\right)\nonumber\\
&&+\gamma _s^2(\varepsilon _s+P_s)[{\bf v}_s({\bf v}_s\cdot {\bf a})-{\bf a}]=0\label{eq26}.
\end{eqnarray}
Although these equations are valid in a FIDO frame, for $\alpha =1$ they reduce to the corresponding special relativistic equations as given by SK \cite{thirty one} which are valid in a frame in which both fluids are at rest. The transformation from the FIDO frame to the commoving (fluid) frame involves a boost velocity, which is simply the freefall velocity onto the black hole, given by
\begin{equation}
v_{\rm ff}=(1-\alpha ^2)^{\frac{1}{2}}\label{eq27}.
\end{equation}
Then the relativistic Lorentz factor becomes $\gamma _{\rm boost}\equiv (1-v_{\rm ff}^2)^{1/2}=1/\alpha $.

\section{Radial Wave Propagation}\label{sec4}
The two-fluid equations describing transverse and longitudinal waves can be separated by considering one dimensional wave propagating in radial $z$ direction and introducing the following complex variables
\begin{eqnarray}
&&v_{sz}(z,t)=u_s(z,t),\hspace{.3cm}v_s(z,t)=v_{sx}(z,t)+{\rm i}v_{sy}(z,t),\nonumber\\
&&B(z,t)=B_x(z,t)+{\rm i}B_y(z,t),\nonumber \\
&&E(z,t)=E_x(z,t)+{\rm i}E_y(z,t)\label{eq28}.
\end{eqnarray}
Then we have
\begin{eqnarray}
v_{sx}B_y-v_{sy}B_x&=&\frac{\rm i}{2}(v_sB^\ast -v_s^\ast B),\nonumber\\
v_{sx}E_y-v_{sy}E_x&=&\frac{\rm i}{2}(v_sE^\ast -v_s^\ast E)\label{eq29}.
\end{eqnarray}
where the $\ast $ denotes the complex conjugate. The continuity equation (\ref{eq12}) and Poisson's equation (\ref{eq21}) become
\begin{eqnarray}
&&\frac{\partial }{\partial t}(\gamma _sn_s)+\frac{\partial }{\partial z}(\alpha \gamma _sn_su_s)=0,\label{eq30}\\
&&\frac{\partial E_z}{\partial z}=4\pi (q_1n_1\gamma _1+q_2n_2\gamma _2).\label{eq31}
\end{eqnarray}
The ${\bf e}_{\hat x}$ and ${\bf e}_{\hat y}$ components of Eqs. (\ref{eq22}) and (\ref{eq23}) give
\begin{eqnarray}
&&\frac{1}{\alpha }\frac{\partial B}{\partial t}=-{\rm i}\left(\frac{\partial }{\partial z}-a\right)E,\label{eq32}\\
&&{\rm i}\frac{\partial E}{\partial t}=-\alpha \left(\frac{\partial }{\partial z}-a\right)B-{\rm i}4\pi e\alpha (\gamma _2n_2v_2-\gamma _1n_1v_1)\quad\quad\label{eq33}.
\end{eqnarray}
Differentiating Eq. (\ref{eq33}) with respect to $t$ and using Eq. (\ref{eq32}), we obtain
\begin{eqnarray}
\left(\alpha ^2\frac{\partial ^2}{\partial z^2}+3\alpha \frac{\partial \alpha}{\partial z}\frac{\partial }{\partial z}-\frac{\partial ^2}{\partial t^2}+(\frac{\partial \alpha }{\partial z})^2\right)E\nonumber\\
=4\pi e\alpha \frac{\partial }{\partial t}(n_2\gamma _2v_2-n_1\gamma _1v_1)\label{eq34}.
\end{eqnarray}
From the ${\bf e}_{\hat x}$ and ${\bf e}_{\hat y}$ components of the equation (\ref{eq26}), the transverse component of the momentum conservation equation is
\begin{eqnarray}
\rho _s\frac{Dv_s}{D\tau }=q_sn_s\gamma _s(E-{\rm i}v_sB_z+{\rm i}u_sB)-u_sv_s\rho _sa\nonumber\\
-v_s\left(q_sn_s\gamma _s{\bf E}\cdot {\bf v}_s+\frac{1}{\alpha }\frac{\partial P_s}{\partial t}\right),\label{eq35}
\end{eqnarray}
where \quad ${\bf E}\cdot {\bf v}_s=\frac{1}{2}(Ev_s^\ast +E^\ast v_s)+E_zu_s$,
and $\rho _s$ is the total energy density defined by
\begin{equation}
\rho _s=\gamma _s^2(\varepsilon _s+P_s)=\gamma _s^2(m_sn_s+\Gamma _gP_s),\label{eq36}
\end{equation}
with $\Gamma _g=\gamma _g/(\gamma _g-1)$.

\section{Linearization and Rindler\\ Approximation}\label{sec5}

In this section we linearized the two-fluid equations for the transverse waves using the following definitions as BHT
\begin{eqnarray}
u_s(z,t)&=&u_{0s}(z)+\delta u_s(z,t),\nonumber\\
n_s(z,t)&=&n_{0s}(z)+\delta n_s(z,t),\nonumber\\
P_s(z,t)&=&P_{0s}(z)+\delta P_s(z,t),\hspace{.2cm}v_s(z,t)=\delta v_s(z,t),\nonumber\\
\rho _s(z,t)&=&\rho _{0s}(z)+\delta \rho _s(z,t),\hspace{.2cm}{\bf E}(z,t)=\delta {\bf E}(z,t),\nonumber\\
{\bf B}_z(z,t)&=&{\bf B}_0(z)+\delta {\bf B}_z(z,t),\hspace{.2cm}{\bf B}(z,t)=\delta {\bf B}(z,t).\quad\label{eq37}
\end{eqnarray}
Here, magnetic field has been chosen to lie along the radial ${\bf e}_{\hat z}$ direction. The relativistic Lorentz factor is also linearized such that
\begin{eqnarray}
&&\gamma _s=\gamma _{0s}+\delta \gamma _s,\mbox{where}\quad\nonumber\\
&&\gamma _{0s}=\left(1-{\bf u}_{0s}^2\right)^{-\frac{1}{2}},\delta \gamma _s=\gamma _{0s}^3{\bf u}_{0s}\cdot \delta {\bf u}_s\label{eq38}.
\end{eqnarray}
The unperturbed radial velocity near the event horizon for each species as measured by a FIDO along ${\bf e}_{\hat z}$ is assumed to be the freefall velocity so that
\begin{equation}
u_{0s}(z)=v_{\rm ff}(z)=[1-\alpha ^2(z)]^{\frac{1}{2}}\label{eq39}.
\end{equation}
The conservation of entropy, Eq. (\ref{eq15}) becomes
\begin{equation}
\delta P_s=\frac{\gamma _gP_{0s}}{n_{0s}}\delta n_s,\label{eq40}
\end{equation}
and from the total energy density, Eq. (\ref{eq36}) redice to
\begin{equation}
\delta \rho _s=\frac{\rho _{0s}}{n_{0s}}\left(1+\frac{\gamma _{0s}^2\gamma _gP_{0s}}{\rho _{0s}}\right)\delta n_s+2u_{0s}\gamma _{0s}^2\rho _{0s}\delta u_s,\label{eq41}
\end{equation}
where $\rho _{0s}=\gamma _{0s}^2(m_sn_{0s}+\Gamma _gP_{0s})$. Using Eq. \ref{eq37} the transverse part of the momentum conservation equation is linearized to
\begin{eqnarray}
\left(\alpha u_{0s}\frac{\partial }{\partial z}+\frac{\partial }{\partial t}-u_{0s}\frac{\partial \alpha }{\partial z}+\frac{{\rm i}\alpha q_s\gamma _{0s}n_{0s}B_0}{\rho _{0s}}\right)\frac{\partial \delta v_s}{\partial t}\nonumber\\
\qquad-\frac{\alpha q_s\gamma _{0s}n_{0s}}{\rho _{0s}}\left(\alpha u_{0s}\frac{\partial }{\partial z}+\frac{\partial }{\partial t}+u_{0s}\frac{\partial \alpha }{\partial z}\right)\delta E=0\label{eq42}.
\end{eqnarray}
Poisson's equation (\ref{eq31}) and Eq. (\ref{eq34}) are linearized to obtain, respectively
\begin{eqnarray}
&&\frac{\partial \delta E_z}{\partial z}=4\pi e(n_{02}\gamma _{02}-n_{01}\gamma _{01})+4\pi e(\gamma _{02}\delta n_2-\gamma _{01}\delta n_1)\nonumber\\
&&+4\pi e(n_{02}u_{02}\gamma _{02}^3\delta u_2-n_{01}u_{01}\gamma _{01}^3\delta u_1),\label{eq43}
\end{eqnarray}
\begin{eqnarray}
\left(\alpha ^2\frac{\partial ^2}{\partial z^2}+3\alpha \frac{\partial \alpha}{\partial z}\frac{\partial }{\partial z}-\frac{\partial ^2}{\partial t^2}+(\frac{\partial \alpha }{\partial z})^2\right)\delta E\nonumber\\
=4\pi e\alpha \left(n_{02}\gamma _{02}\frac{\partial \delta v_2}{\partial t}-n_{01}\gamma _{01}\frac{\partial \delta v_1}{\partial t}\right)\label{eq44}.
\end{eqnarray}
The Rindler coordinate system, in which space is locally Cartesian, provides a good approximation to the SAdS metric near the event horizon. In this coordinates the two-fluid equations can be solved just inside and outside the event horizon and cannot be used to the case for extremal black holes because there are no Rindler coordinates locally near horizons. Near the event horizon of SAdS black hole, choosing the coordinate
\begin{equation}
f\approx (r-r_+)\times 2\Big(\frac{(r_+/\ell)^2+M\ell/r^2_+}{r_+}\Big)=(\kappa z)^2,\label{eq45}
\end{equation}
we may write the matric (\ref{eq1}) approximately as
\begin{equation}
ds^2=-(\kappa z)^2dt^2+\frac{(\ell \kappa )^2}{(r_+/\ell+M\ell/r^2_+)^2}dz^2+r^2(d\theta ^2+{\rm sin}^2\theta d\varphi ^2).\label{eq46}
\end{equation}
This matric becomes a Rindler spacetime with the surface gravity
\begin{equation}
\kappa =\frac{1}{\ell}\Big(\frac{r_+}{\ell}+\frac{M\ell}{r^2_+}\Big).\label{eq47}
\end{equation}
In the Rindler coordinates the SAdS spacetime becomes
\begin{equation}
ds^2=-\alpha^2dt^2+dx^2+dy^2+dz^2,\label{eq48}
\end{equation}
where
\begin{equation}
x=r_+\left(\theta -\frac{\pi }{2}\right),\hspace{.5cm}y=r_+\varphi ,\hspace{.5cm}z=\frac{\alpha }{\kappa}\label{eq49}.
\end{equation}
The standard lapse function in Rindler coordinates becomes $\alpha =z\kappa$.

\section{Radial Dependence of\\ Equilibrium Parameters}\label{sec6}

The dependence of the equilibrium field and fluid parameters on the radial coordinate for transverse waves is discussed here. The equilibrium parameters are expressed in terms of limiting horizon values and then for the lapse function and the equilibrium fields and fluid quantities a local (or mean-field) approximation is used.
From the continuity equation (\ref{eq30}), we have
\[
r^2\alpha \gamma _{0s}n_{0s}u_{0s}=\mbox{const.}=r_+^2\alpha _+\gamma _+n_+u_+,
\]
where the subscript $+$ indicate the limiting values at the event horizon. The freefall velocity at the horizon becomes unity so that $u_+=1$. Since $u_{0s}=v_{\rm ff}$, $\gamma _{0s}=1/\alpha $; and hence $\alpha \gamma _{0s}=\alpha _+\gamma _+=1$. Also, because $v_{\rm ff}=(r_+/r)^{1/2}\zeta $ with $0<\zeta\leq 1$. The value for $\zeta=1$ corresponds to the Schwarzschild black hole. The number density for each species can be written as follows:
\begin{equation}
n_{0s}(z)=n_{+s}v_{\rm ff}^3(z)/\zeta ^4\label{eq50},
\end{equation}
The equation of state Eq. (\ref{eq15}) lead to write the unperturbed pressure,
\begin{equation}
P_{0s}(z)=P_{+s}(\frac{n_{0s}}{n_{+s}})^{\gamma _g}\label{eq51}.
\end{equation}
In terms of the freefall velocity, it can be written as
\begin{equation}
P_{0s}(z)=P_{+s}v_{\rm ff}^{3\gamma _g}(z)/\zeta ^{16/3}\label{eq52}.
\end{equation}
Since $P_{0s}=k_Bn_{0s}T_{0s}$, then with $k_B=1$, the temperature profile is
\begin{equation}
T_{0s}=T_{+s}v_{\rm ff}^{3(\gamma _g-1)}(z)/\zeta ^{4/3}\label{eq53}.
\end{equation}
The unperturbed magnetic field is purely in the radial direction. It does not experience effects of spatial curvature. From the flux conservation equation $\nabla \cdot {\bf B}_0=0$ we have
\[
r^2B_0(r)=\mbox{const.},
\]
in terms of freefall velocity one can obtains the unperturbed magnetic field in the form
\begin{equation}
B_0(z)=B_+v_{\rm ff}^4(z)/\zeta ^4,\label{eq54}
\end{equation}
where $v_{\rm ff}=[1-\alpha ^2(z)]^{1/2}$.

In this work our main subject is how the waves behave in the vicinity of black hole and how the horizon affects these waves in anti-de sitter spacetime. Since the plasma is situated relatively close to the horizon, $\alpha ^2\ll 1$, then a relatively small change in distance $z$ will make a significant difference to the magnitude of $\alpha $. We need to choose a sufficiently small range in $z$ for which the values of $\alpha $ does not vary much. So we consider thin layers in the ${\bf e}_{\hat z}$ direction, each layer with its own $\alpha _0$, where $\alpha _0$ is some mean value of $\alpha $ within a particular layer. Then a more complete picture can be built up by considering a large number of layers within a chosen range of $\alpha _0$ values.

The local approximation imposes the restriction that the wavelength must be smaller in magnitude than the scale of the gradient of the lapse function $\alpha $, {\it i.e.},
\[
\lambda <\left(\frac{\partial \alpha }{\partial z}\right)^{-1}=2r_+\simeq 5.896\zeta \times 10^5{\rm cm},
\]
or, equivalently,
\[
k>\frac{2\pi }{2r_+}\zeta^{-1} \simeq 1.067\zeta^{-1}\times 10^{-5}{\rm cm}^{-1},0<\zeta \leq 1,
\]
for a black hole of mass $\sim 1M_\odot $.

In the local approximation for $\alpha $, $\alpha \simeq \alpha _0$ is valid within a particular layer. Hence, the unperturbed fields and fluid quantities and their derivatives, which are functions of $\alpha $, take on their corresponding \lq \lq mean-field\rq \rq values for a given $\alpha _0$. Then the coefficients in Eqs. (\ref{eq48}), (\ref{eq41}), and (\ref{eq52}) are constants within each layer with respect to $\alpha $ (and therefore $z$ as well). So it is possible to Fourier transform the equations with respect to $z$, using plane-wave-type solutions for the perturbations of the form $\sim e^{i(kz-\omega t)}$ for each $\alpha _0$ layer.\\
Using Fourier transformation, Eqs. (\ref{eq51}) and (\ref{eq53}) become
\begin{equation}
\delta E=\frac{{\rm i}4\pi e\alpha _0\omega (n_{02}\gamma _{02}\delta v_2-n_{01}\gamma _{01}\delta v_1)}{\alpha _0k(\alpha _0k-{\rm i}3\kappa)-\omega ^2-\kappa^2},\label{eq55}
\end{equation}
\begin{eqnarray}
&&\omega \left(\alpha _0ku_{0s}-\omega +{\rm i}u_{0s}\kappa+\frac{\alpha _0q_s\gamma _{0s}n_{0s}B_0}{\rho _{0s}}\right)\delta v_s\nonumber\\
&&\qquad-{\rm i}\alpha _0\frac{q_s\gamma _{0s}n_{0s}}{\rho _{0s}}\left(\alpha _0ku_{0s}-\omega -{\rm i}u_{0s}\kappa \right)\delta E=0\nonumber\\
\label{eq56}.
\end{eqnarray}

\section{Dispersion Relation and\\ Numerical Solution of Modes}\label{sec7}
The dispersion relation for the transverse electromagnetic wave modes may be put
using Eqs. (\ref{eq55}) and (\ref{eq56}) in the form
\begin{eqnarray}
&&\left[K_\pm \left(K_\pm \pm {\rm i}\kappa \right)-\omega ^2+\kappa^2\right]=\alpha _0^2\Bigg\{\frac{\omega _{p1}^2(\omega -u_{01}K_\pm )}{(u_{01}K_\mp -\omega -\alpha _0\omega _{c1})}\nonumber\\
&&+\frac{\omega _{p2}^2(\omega -u_{02}K_\pm )}{(u_{02}K_\mp -\omega +\alpha _0\omega _{c2})}\Bigg\}\label{eq57}
\end{eqnarray}
for either electron-positron or electron-ion plasma.

    Here the local plasma frequency for transverse wave $\omega _{ps}=\sqrt{{4\pi e^2\gamma _{0s}^2n_{0s}^2}/\rho _{0s}}$, depends on the local number density of electrons and also upon the local value of the lapse function $\alpha $, and the  local cyclotron frequency
 $\omega _{cs}={e^2\gamma _{0s}n_{0s}B_0}/{\rho_{0s}}$, depends upon both the local magnetic field and the lapse function, with $K_{\pm }=\alpha_0 k\pm {i}/{2r_+}$. The $+$ and $-$ denote the left (L) and right (R) modes, respectively. The dispersion relation for the L mode is obtained by taking the complex conjugate of the dispersion relation for the R mode. The two modes have the same dispersion relation in the special relativistic case.\\
The dispersion relation we have derived here for transverse waves are complicated enough and an analytical solution is cumbersome and unprofitable, even in the simplest cases for the electron-positron plasma where both species are assumed to have the same equilibrium parameters. We solve numerically the dispersion relation in the form of a matrix equation as follows:
\begin{equation}
(A-kI)X=0.\label{eq58}
\end{equation}
The eigenvalue is chosen to be the wave number $k$, the eigenvector $X$ is given by the relevant set of perturbations, and $I$ is the identity matrix.
We need to write the perturbation equations in an appropriate form. We introduce the following set of dimensionless variables:
\begin{eqnarray}
&&\tilde \omega =\frac{\omega }{\alpha _0\omega _\ast },\quad \tilde k=\frac{kc}{\omega _\ast },\quad k_+=\frac{\kappa}{\omega _\ast },\nonumber\\
&&\delta \tilde u_s=\frac{\delta u_s}{u_{0s}},\quad \tilde v_s=\frac{\delta v_s}{u_{0s}},\quad \delta \tilde n_s=\frac{\delta n_s}{n_{0s}},\nonumber\\
&&\delta \tilde B=\frac{\delta B}{B_0},\quad \tilde E=\frac{\delta E}{B_0},\quad \delta \tilde E_z=\frac{\delta E_z}{B_0}.\label{eq59}
\end{eqnarray}

For an electron-positron plasma, $\omega _{p1}=\omega _{p2}$ and $\omega _{c1}=\omega _{c2}$; hence, $\omega _\ast $ is defined as
\begin{eqnarray*}
\omega _\ast =\left\{\begin{array}{rl}&\omega _c\hspace{2.8cm}\mbox{Alfv\'en modes},\\
&\\
&(2\omega _p^2+\omega _c^2)^{1/2}\qquad
\mbox{high frequency modes},\end{array}\right.
\end{eqnarray*}
where $\omega _p=\sqrt{\omega _{p1}\omega _{p2}}$ and $\omega _c=\sqrt{\omega _{c1}\omega _{c2}}$. However, for the case of an electron-ion plasma, the plasma frequency and the cyclotron frequency are different for each fluid, and so the choice of $\omega _\ast $ is a more complicated matter. For simplicity, we assume that
\begin{eqnarray*}
\omega _\ast =\left\{\begin{array}{rl}&\frac{1}{\sqrt{2}}(\omega _{c1}^2+\omega _{c2}^2)^{1/2}\quad \mbox{Alfv\'en modes},\\
&\\
&(\omega _{\ast 1}^2+\omega _{\ast 2}^2)^{1/2}\qquad
\mbox{high frequency modes},\end{array}\right.
\end{eqnarray*}
where $\omega _{\ast s}^2=(2\omega _{ps}^2+\omega _{cs}^2)$.

The dimensionless eigenvector for the transverse set of equations is
\begin{equation}
\tilde X_{\rm transverse}=\left[\begin{array}{c}\delta \tilde v_1\\\delta \tilde v_2\\\delta \tilde B\\\delta \tilde E\end{array}\right]\label{eq60}.
\end{equation}

When Eqs. (\ref{eq32}) and (\ref{eq33}) are linearized and Fourier transformed, they take the forms
\begin{eqnarray}
&&\left(k-\frac{{\rm i}\kappa}{\alpha _0}\right)\delta E+\frac{{\rm i}\omega }{\alpha _0}\delta B=0,\label{eq61}\\
&&\frac{{\rm i}\omega }{\alpha _0}\delta E=\left(k-\frac{{\rm i }\kappa}{\alpha _0}\right)\delta B
+4\pi e(\gamma _{02}n_{02}\delta v_2\nonumber\\
&&-\gamma _{01}n_{01}\delta v_1)\label{eq62}.
\end{eqnarray}
Using Eq. (\ref{eq59}), we write Eq. (\ref{eq56}), (\ref{eq61}), and (\ref{eq62}) in the dimensionless form:
\begin{eqnarray}
\tilde k\delta \tilde v_s&=&\left(\frac{\tilde \omega }{u_{0s}}-\left(\frac{q_s}{e}\right)\frac{\omega _{cs}}{u_{0s}\omega _\ast }-\frac{{\rm i}k_e}{\alpha _0}\right)\delta \tilde v_s\nonumber\\
&&+\left(\frac{q_s}{e}\right)\frac{\omega _{cs}}{u_{0s}\omega _\ast }\delta \tilde B-{\rm i}\left(\frac{q_s}{e}\right)\frac{\omega _{cs}}{u_{0s}\omega _\ast }\delta \tilde E,\label{eq63}\\
\tilde k\delta \tilde E&=&-{\rm i}\tilde \omega \delta \tilde B+\frac{{\rm i}k_e}{\alpha _0}\delta \tilde E,\label{eq64}\\
\tilde k\delta \tilde B&=&u_{01}\frac{\omega _{p1}^2}{\omega _{c1}\omega _\ast }\delta \tilde v_1-u_{02}\frac{\omega _{p2}^2}{\omega _{c2}\omega _\ast }\delta \tilde v_2+\frac{{\rm i}k_e}{\alpha _0}\delta \tilde B+{\rm i}\tilde \omega \delta \tilde E\nonumber\\
\label{eq65}.
\end{eqnarray}
These are the equations in the required form to be used as input to Eq. (\ref{eq57}). Equations (\ref{eq63}), (\ref{eq64}), and (\ref{eq65}) can be written as
$(\tilde A-\tilde kI)\tilde X=0.$  We have calculated all the eigenvalues $\tilde k$ of the complex matrix $\tilde A$ to draw the Alfv\'en and high frequency transverse wave modes using MATLAB.
\newpage
\begin{figure}[t]\label{fig1}
 \begin{center}
 \includegraphics[scale=0.34]{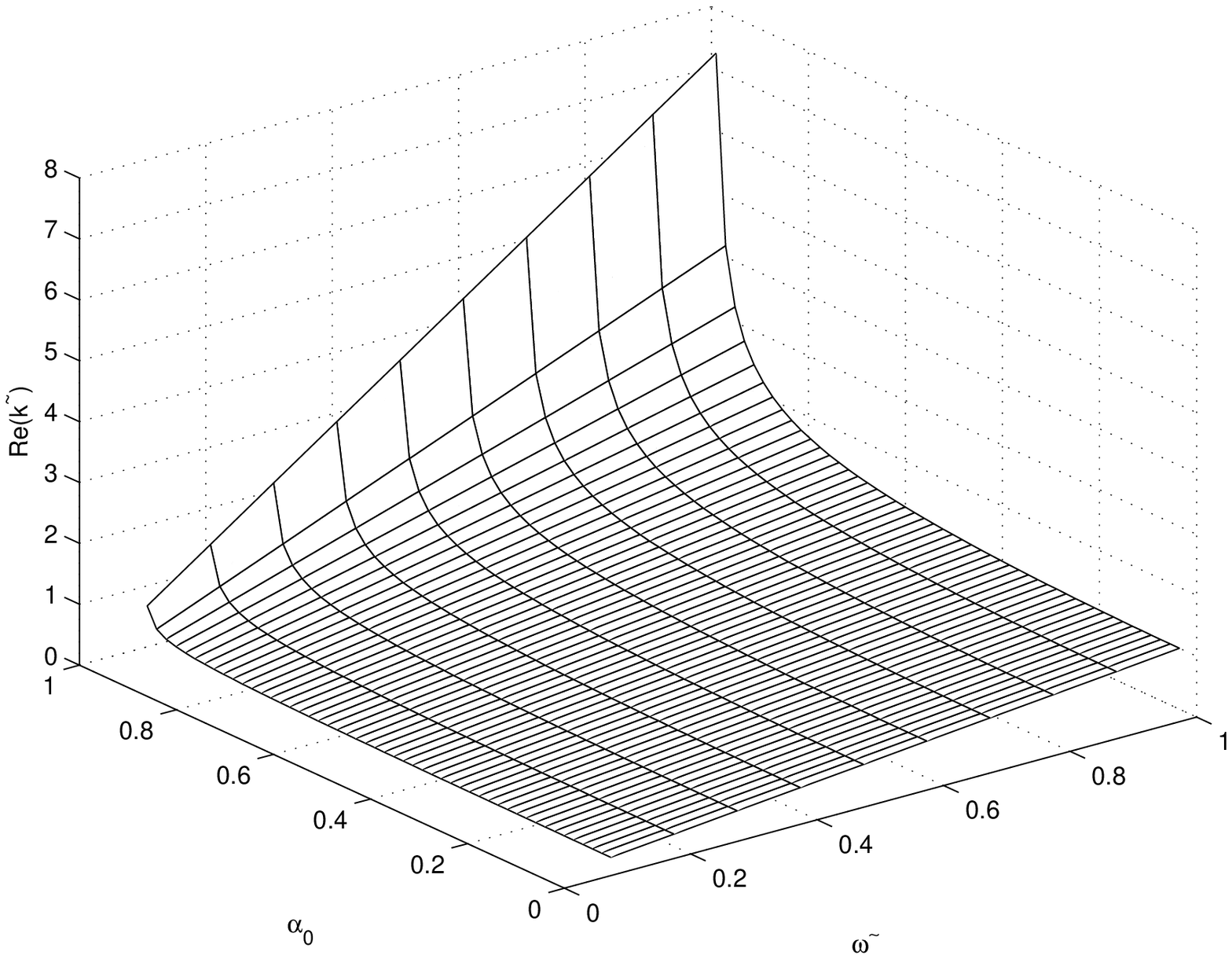}\\
 \includegraphics[scale=0.34]{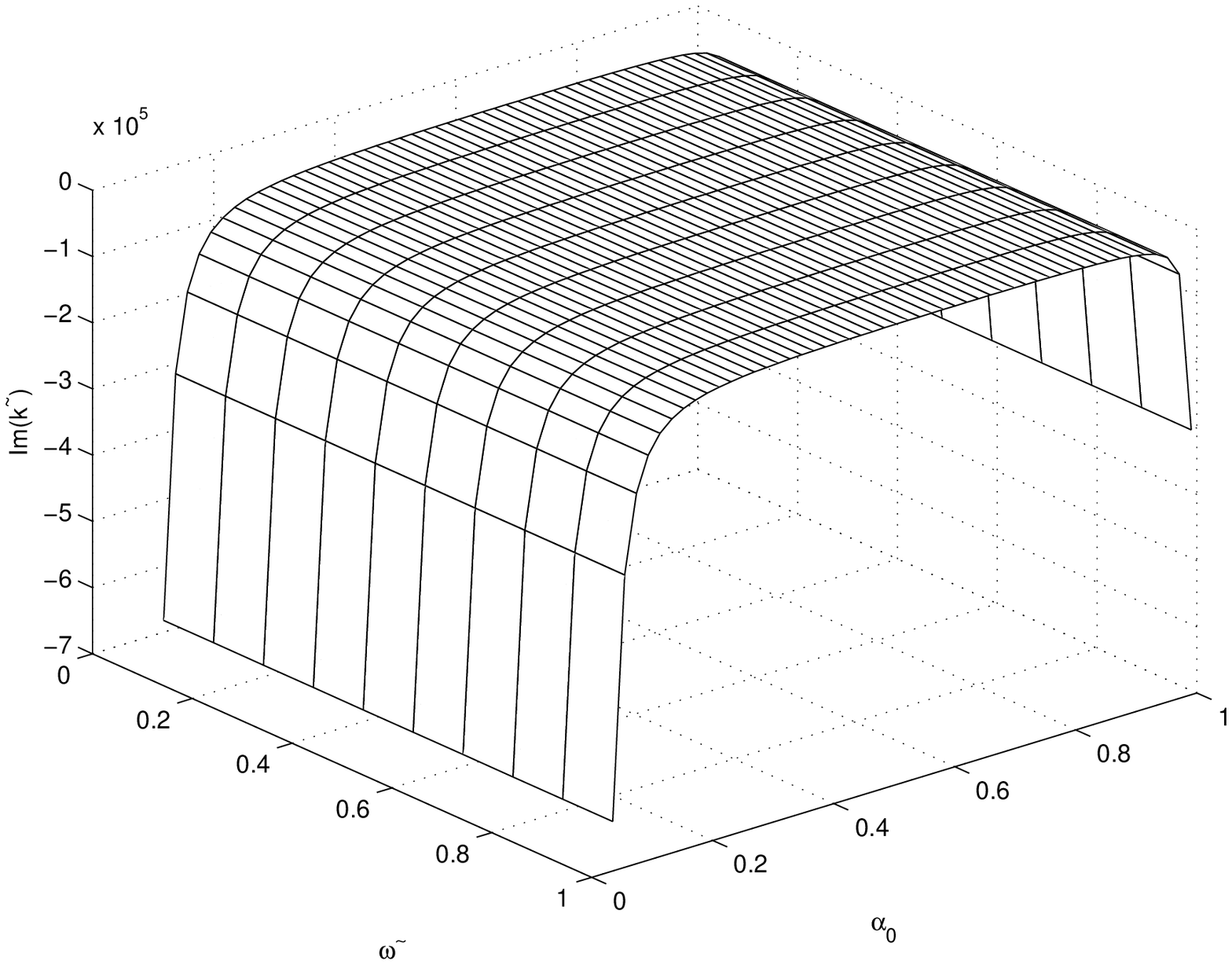}\\
 \includegraphics[scale=0.34]{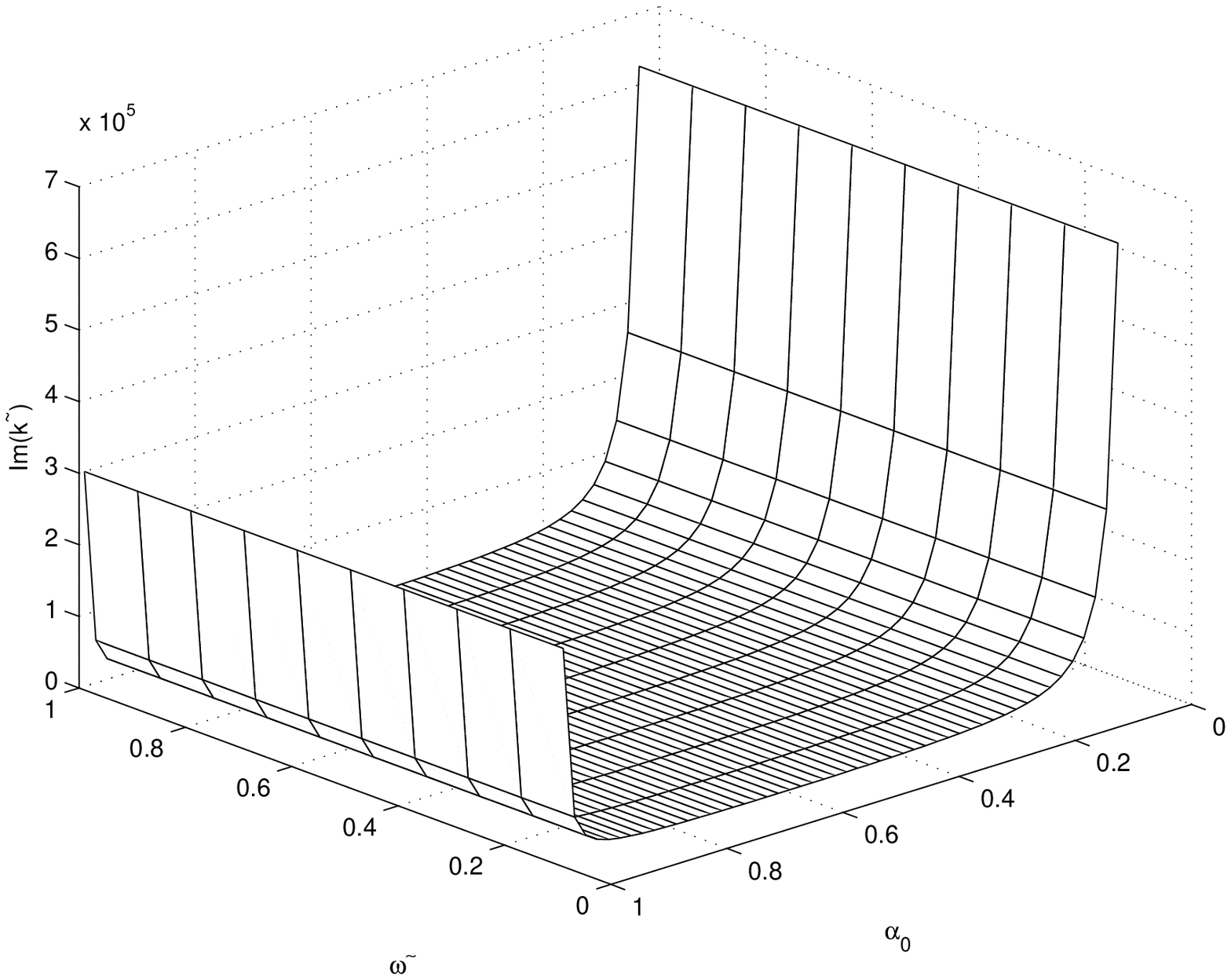}
\end{center}
\caption{\it Top: Real part of the complex conjugate pair of Alfv\'en modes for electron-positron plasma. Center: Imaginary part of growth mode. Bottom: Imaginary part of damped mode}
\end{figure}
\section{Alfv\'en Modes}\label{sec8}
\centerline{\large{\it{\bf 1. Electron-positron Plasma}}}
For the ultrarelativistic electron-positron plasma in the special
relativistic formulation, only one real Alfv\'en mode is found to exist
\cite{thirty one} this is because both the left and right circularly polarized modes were described by the same dispersion relation thereby leading to the same mode and BHT found two Alfv\'en modes for general relativistic formulation close to the horizon of a Schwarzschild black hole. In our present work we have found four modes for electron-positron plasma two of which are shown in Fig. are complex conjugate pair and are similar with the two modes of BHT \cite{six}. The third mode is damped similar to the damped mode of Fig. other two modes are also damping and growth modes and have a little difference in with smaller damping rate and the fourth mode is equivalent to the third mode with opposite in sense and shows growth.
\begin{figure}[h]\label{fig2}
\begin{center}
 \includegraphics[scale=.34]{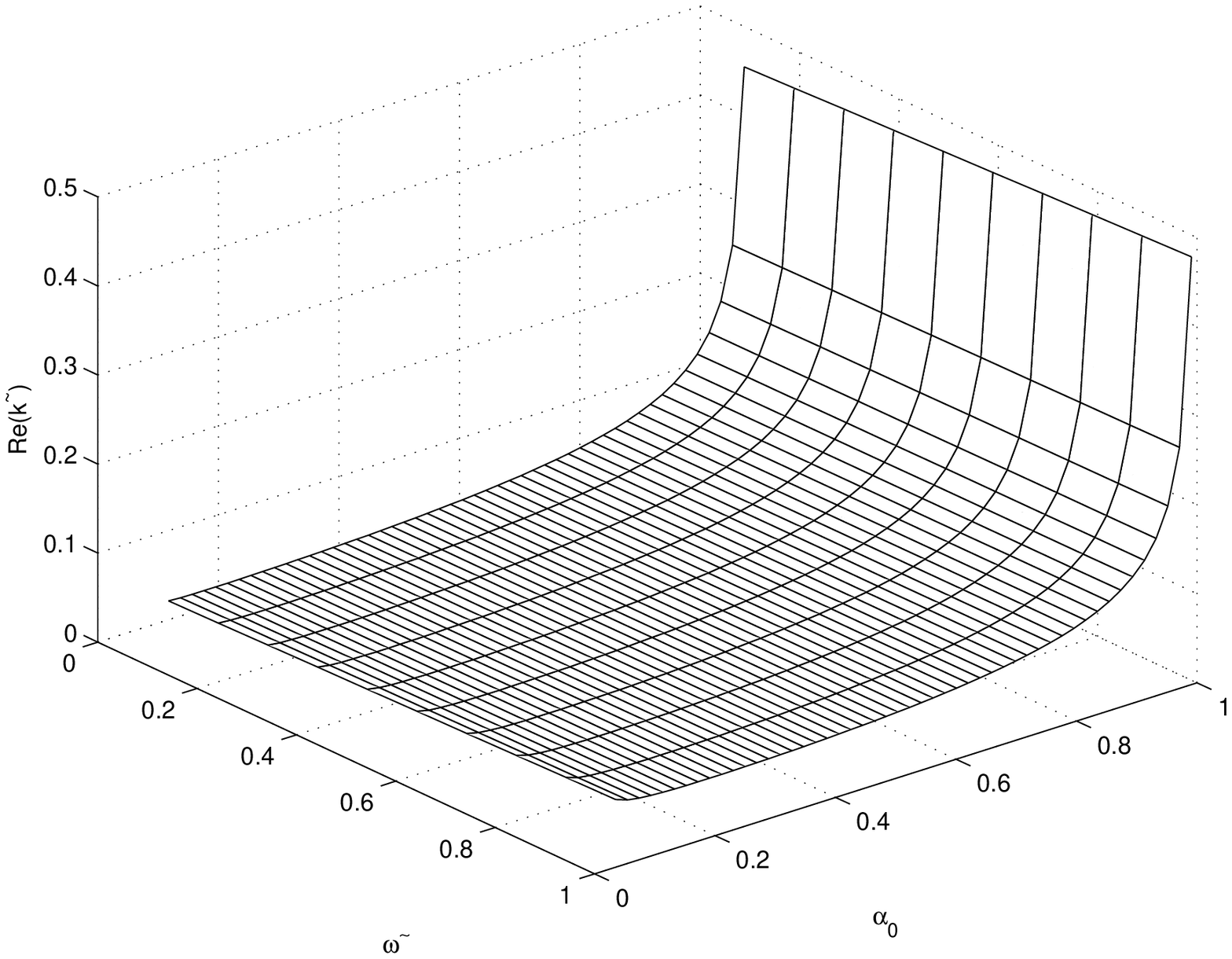}
 \includegraphics[scale=.34]{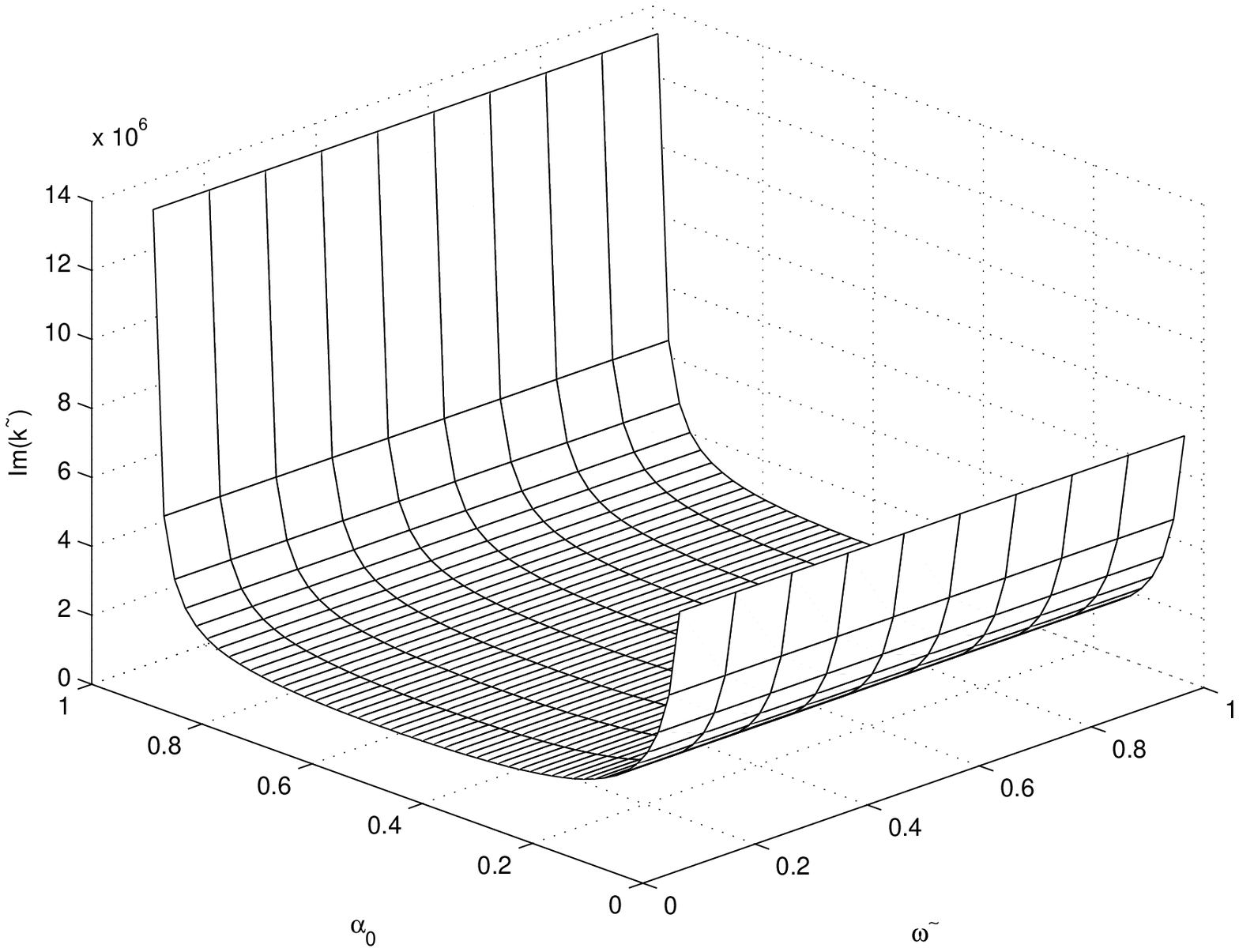}
 \includegraphics[scale=.34]{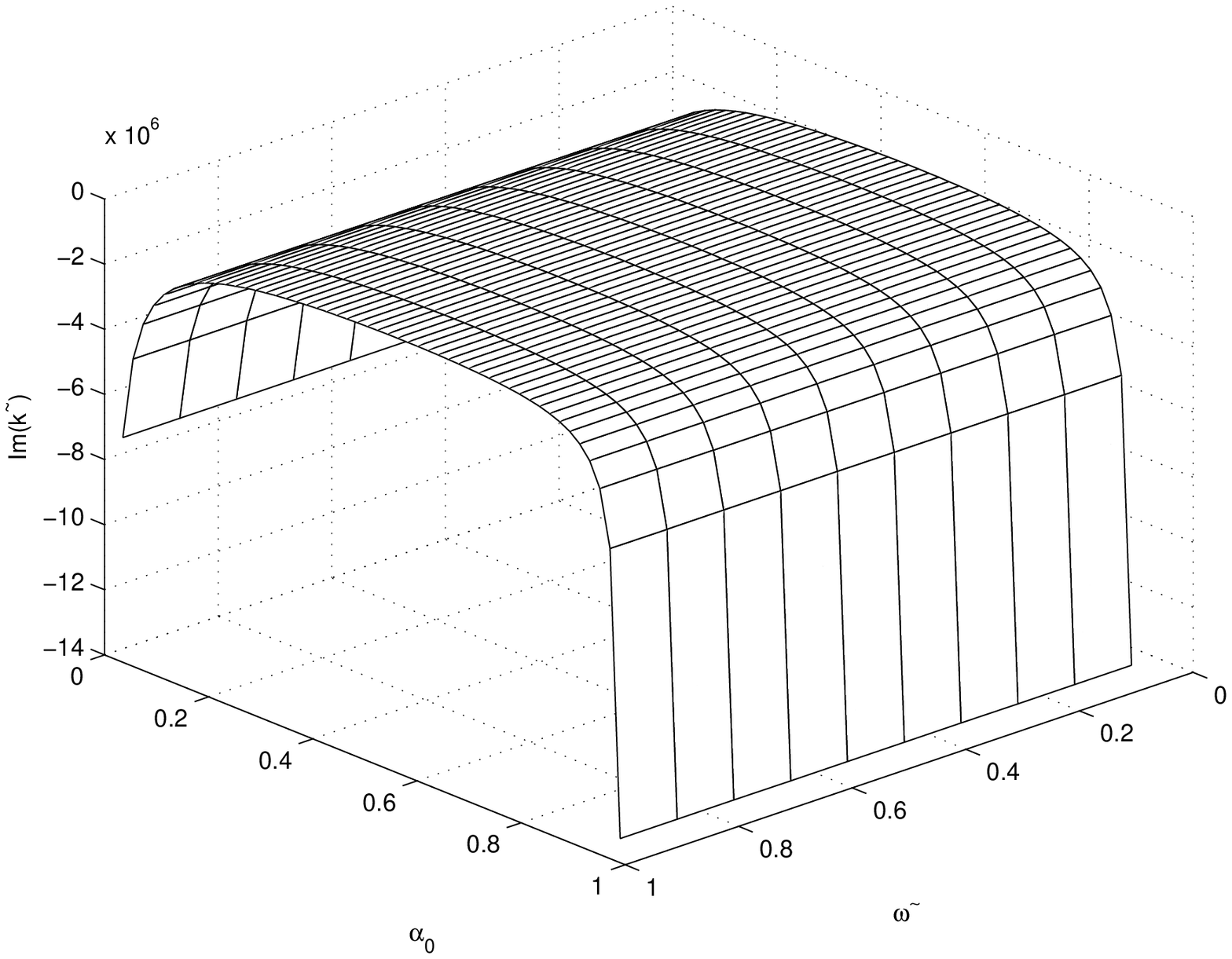}
\end{center}
\caption{\it Top: Real part of the complex conjugate pair of Alfv\'en modes for electron-ion plasma. Center: Imaginary part of damped mode. Bottom: Imaginary part of growth mode}
\end{figure}
 The damping and growth rates of all the modes are smaller by several orders of magnitude in comparisons with the corresponding modes for Schwarzschild black hole as investigated by BHT. These four modes for the electron-positron plasma coalesce with two modes of BHT and a single mode of SK and so yielding BHT and SK results. Here $\rm for Im(k)>0 $ corresponds to damping and $\rm  for Im(k)<0 $ to growth. This is because the convention we have used is $e^{{\rm i}kz}=e^{{\rm i}[{\rm Re}(k)+{\rm iIm}(k)]z}$. The limiting horizon values are taken to be $n_{+s}=10^{18}{\rm cm}^{-3}$,  $T_{+s}=10^{10}{\rm K}$, $B_+=3\times 10^6{\rm G}$, $\gamma _g=4/3$, and $\eta=0.9$ with black hole mass $M=5M_\odot$.
\newpage
\begin{figure}[t]\label{fig3}
\begin{center}
 \includegraphics[scale=.34]{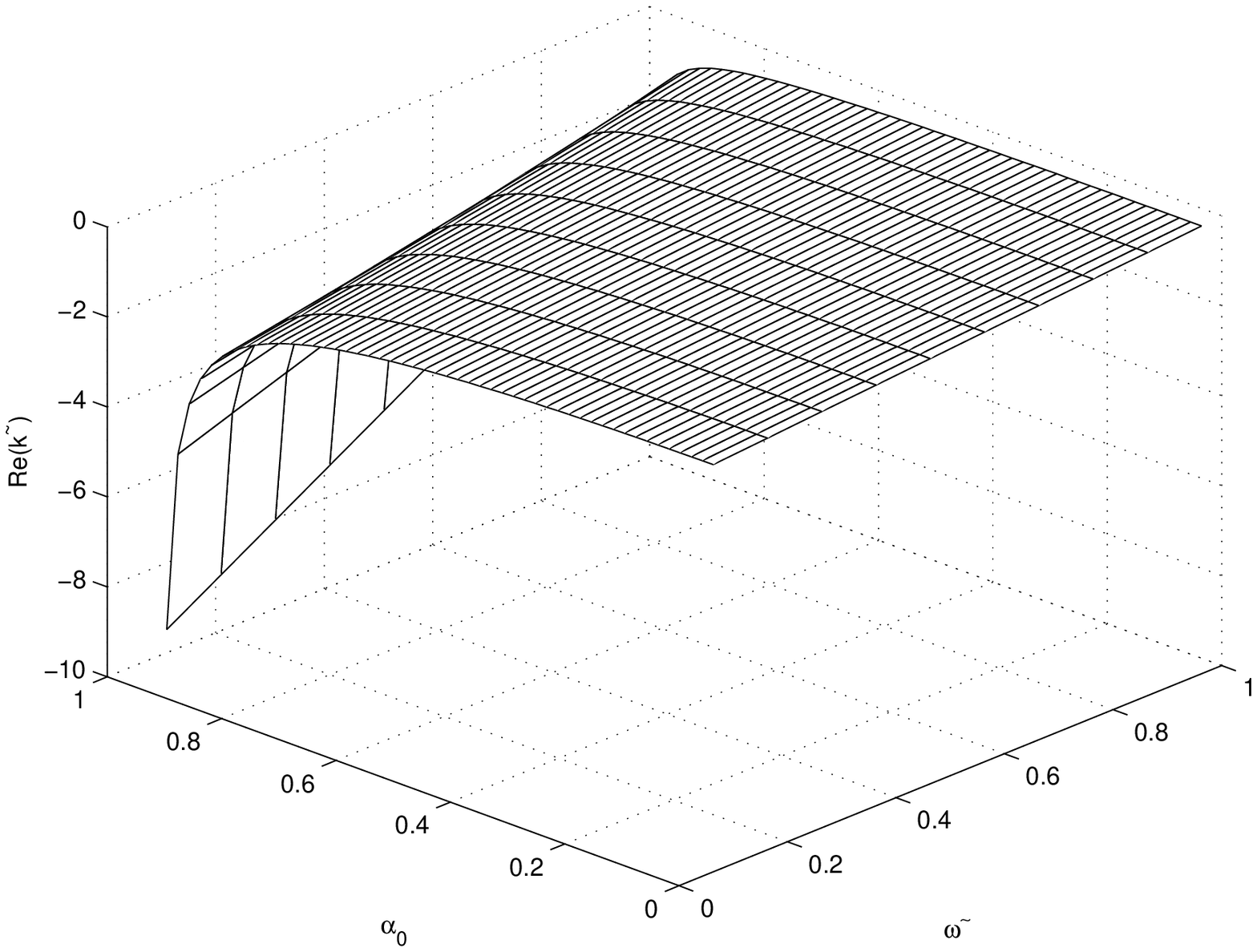}
 \includegraphics[scale=.34]{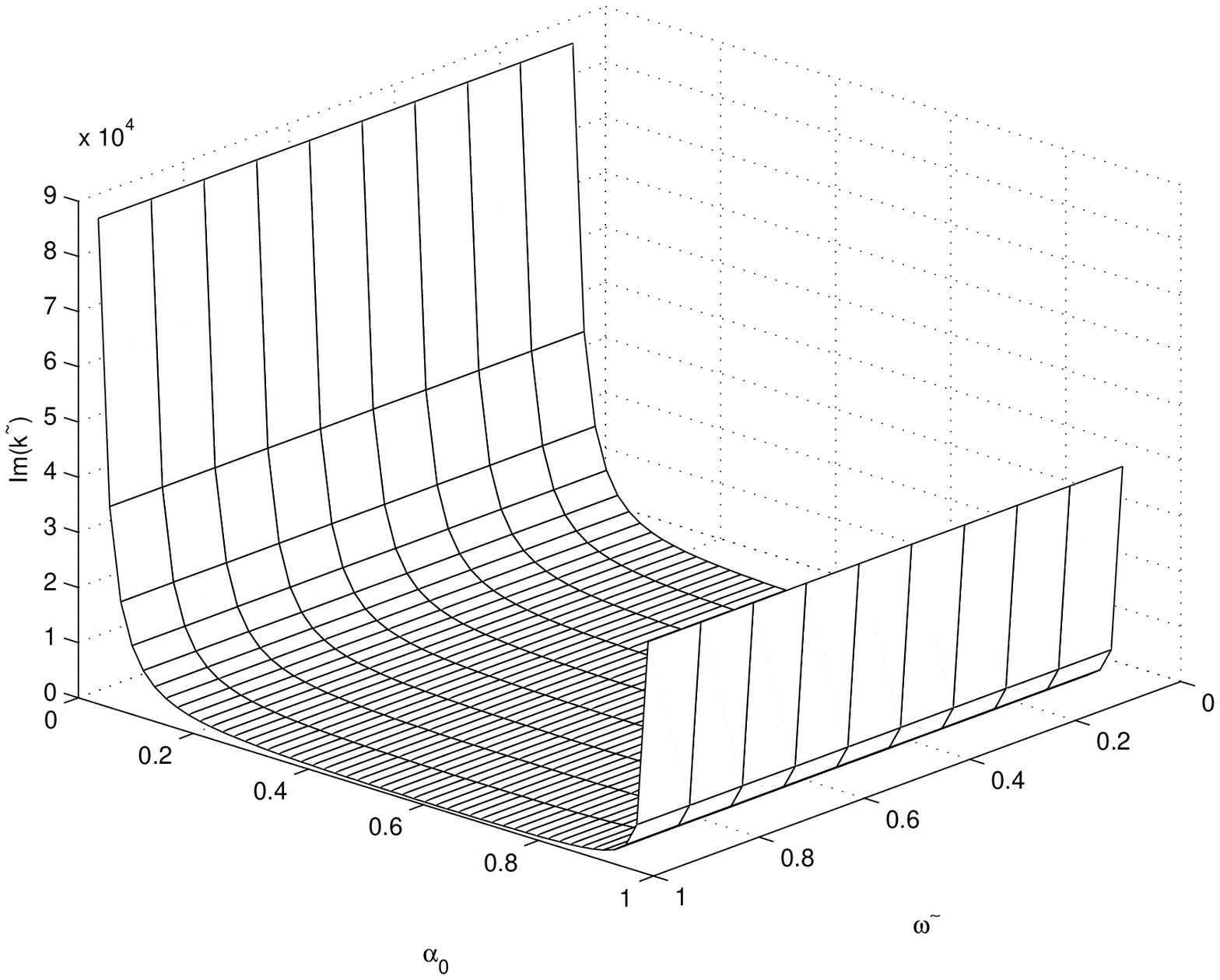}
\end{center}
\caption{\it Top: Real part of Alfv\'en damped mode for the electron-ion plasma. Bottom: Imaginary part of damped mode.}
\end{figure}

\centerline{\large{\it {\bf 2. Electron-ion Plasma}}}
In this case there exist four modes, two of which are shown in Fig. 2 and Fig. 3, are growth and damped respectively and are different modes. The other two modes shown in fig. 4 are equivalent to that of Buzzi et al. \cite{six} and is complex conjugate growth and damping modes. The differences in the magnitudes of the $\omega _{c1}$ and $\omega _{c2}$ apparently lead to take the frequencies from their negative (and therefore unphysical) values for the electron-positron case to positive physical values for the electron-ion case. These changes are thus because of the difference in mass and density factors as between the positrons and ions. These four modes for electron-ion plasma are analogous to the four modes obtained by BHT in the Schwarzschild case. It is evident that the growth and damping rates are independent of the frequency, but depended only on the distance from the black hole horizon through $\alpha _o$. The limiting horizon values for electron-ion plasma are chosen as $n_{+1}=10^{18}{\rm cm}^{-3}$, $T_{+1}=10^{10}{\rm K}$, $n_{+2}=10^{15}{\rm cm}^{-3}$, $T_{+2}=10^{12}{\rm K}$, $\eta=0.9$ , and the mass of the black hole $M=5M_\odot$.
\newpage
\begin{figure}[t]\label{fig4}
\begin{center}
 \includegraphics[scale=.34]{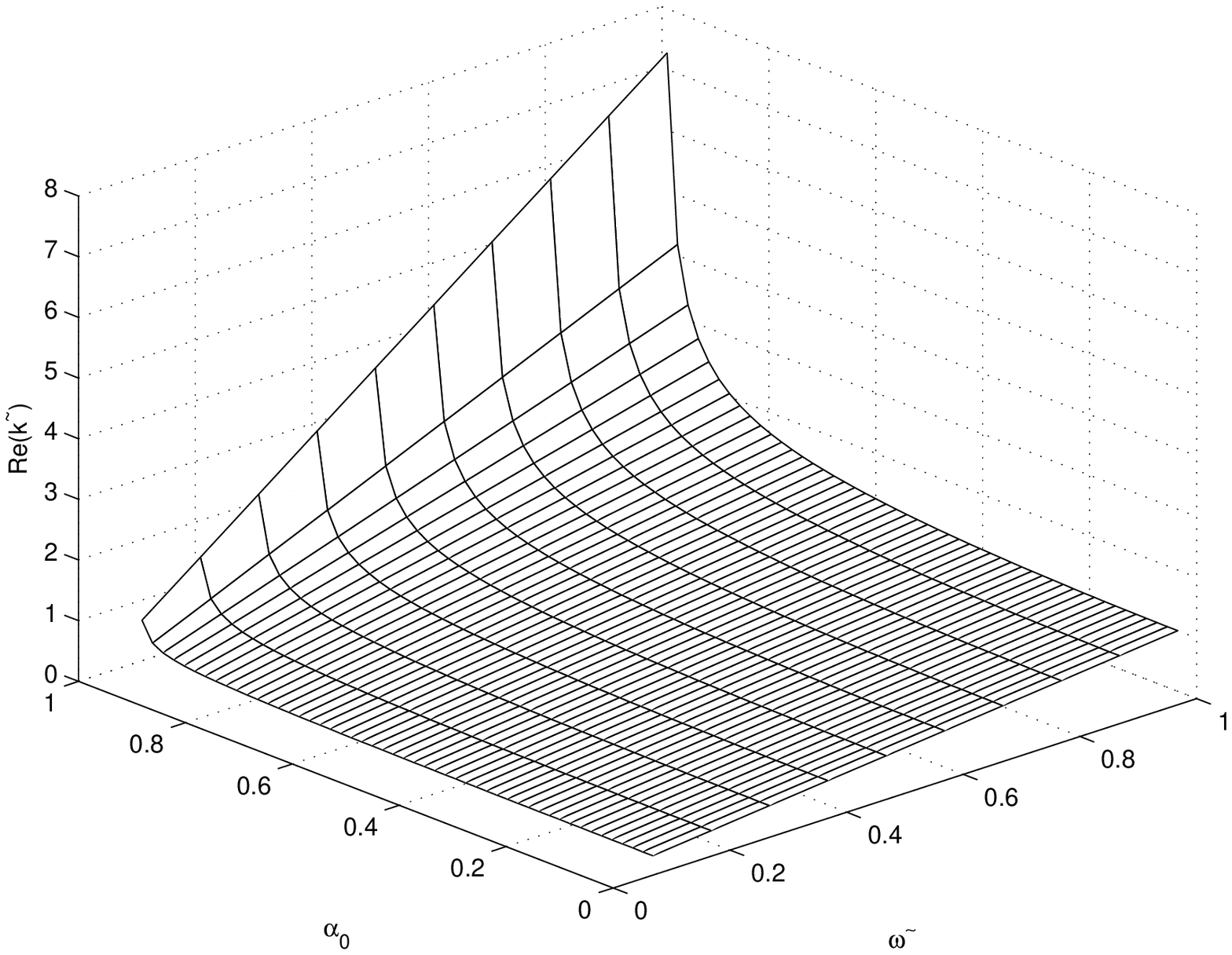}
 \includegraphics[scale=.34]{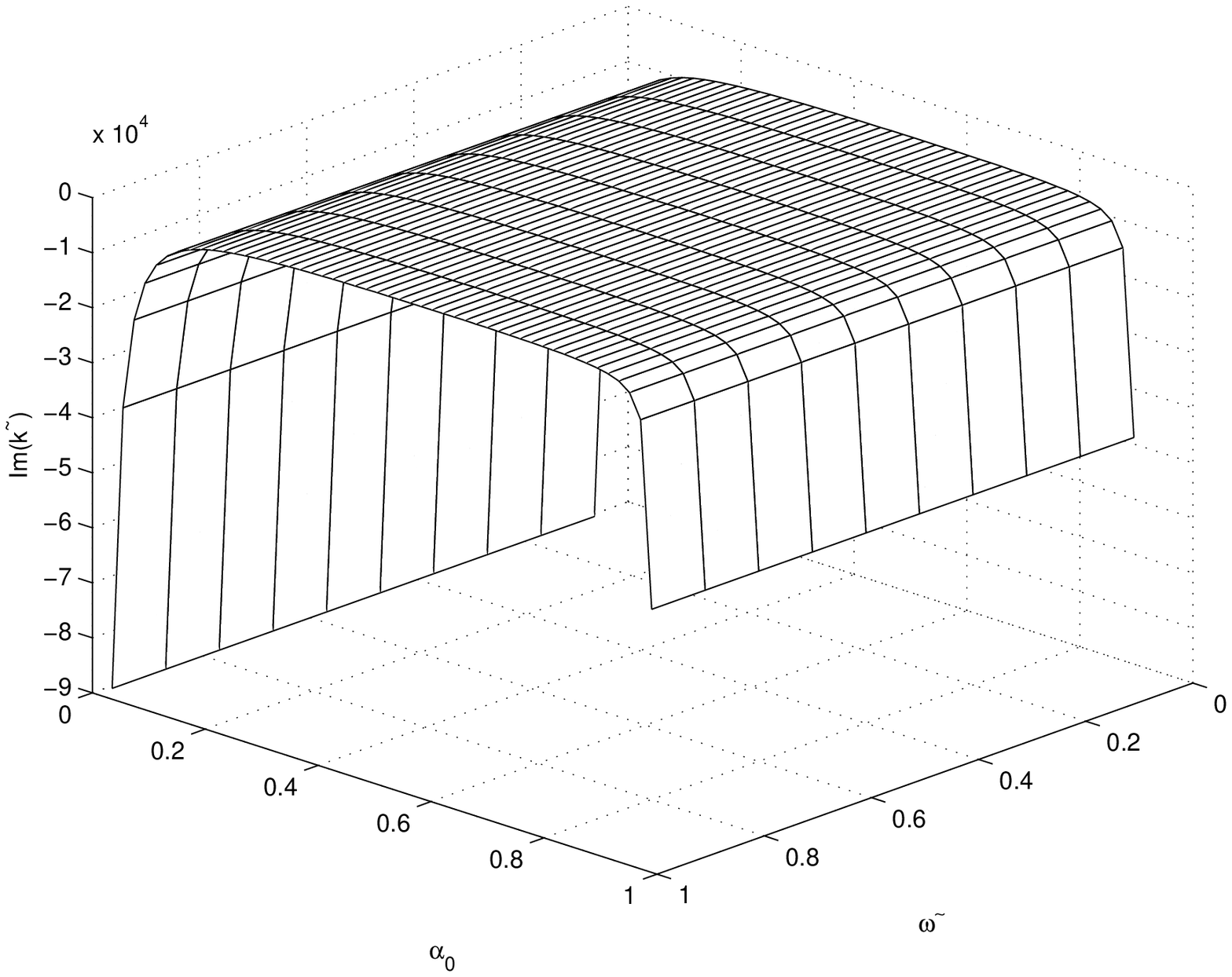}
\end{center}
\caption{\it Top: Real part of Alfv\'en growth mode for the electron-ion plasma. Bottom: Imaginary part of growth mode.}
\end{figure}

\section{High Frequency\\ Transverse Modes}\label{sec9}
\centerline{\large{\it{\bf 1. Electron-positron Plasma}}}
There exist four high frequency electromagnetic modes for the electron-positron plasma. Fig. 5 shows two modes having same real parts equivalent to the real part of Fig.1. The mode shown in the center is completely a growth mode but the bottom mode is damped for higher frequency and $\alpha_o>0.5$, but growth for $\alpha_o<0.5$ and lower frequencies. Thus at a distance from the horizon corresponding to $\alpha _o< 0.5$, it appears that energy is no longer fed into wave mode by the gravitational field but begins to be drained from the waves. The other two modes shown in Fig. 6 are not complex conjugate although the real part is equal. The center mode shown in Fig. 6 is damped for the entire frequency domain but the bottom mode is damped for most of the frequency domain and for $\alpha_o>0.3$ and shows growth for lower frequencies and for $\alpha _o<0.3$. Comparing the center and bottom modes of Figs. 5 and 6 it is quite clear that in the special relativistic limit the mode shown in Figs. 5 and 6 are growing and damped respectively and coalesce into only one purely high frequency mode in the special relativistic case \cite{thirty one} for the ultrarelativistic electron-positron plasma. These four modes reduce to the three modes of Buzzi et al. \cite{six} in the Schwarzschild case.
\newpage
\begin{figure}[h]\label{fig5}
\begin{center}
 \includegraphics[scale=.34]{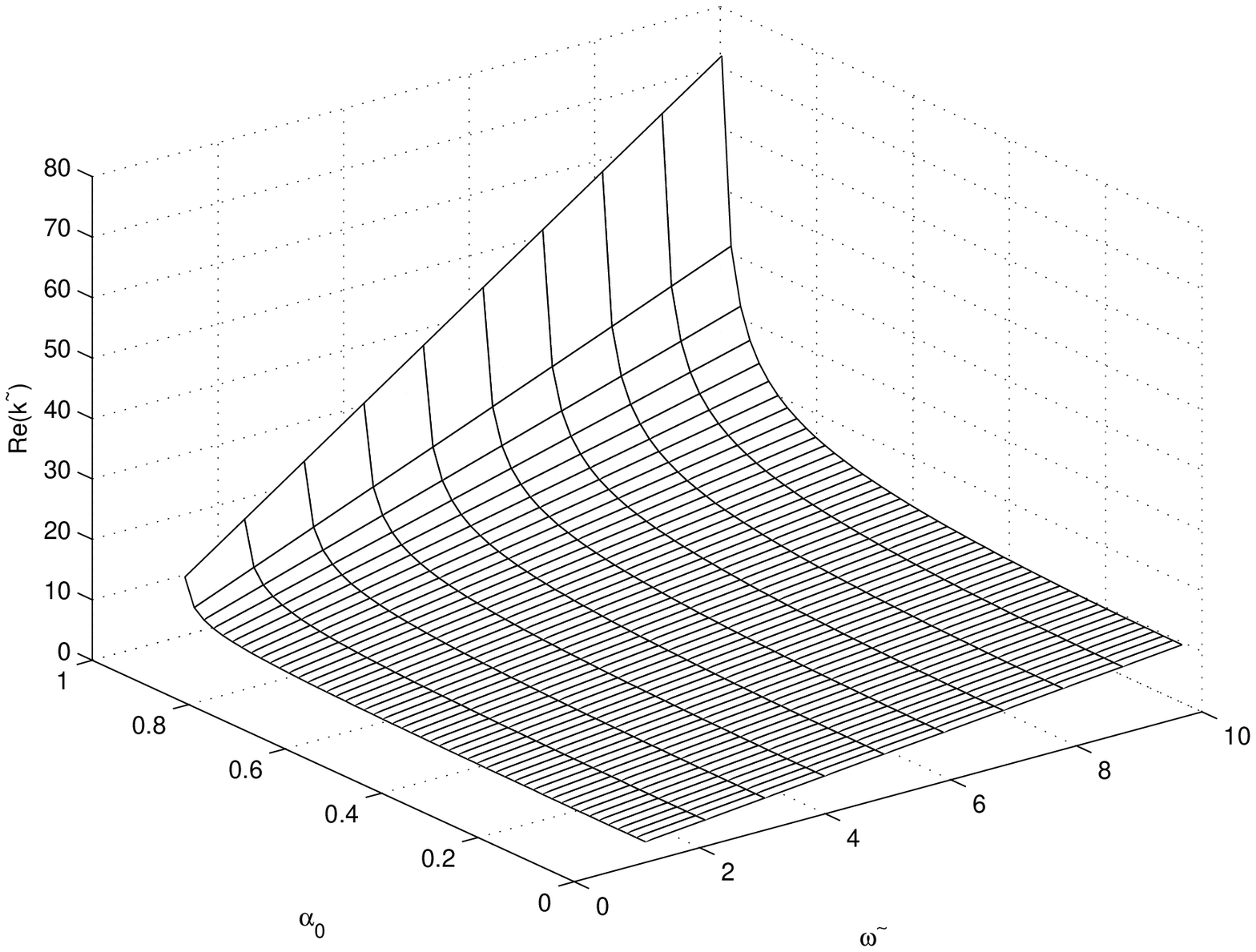}\\
 \includegraphics[scale=.34]{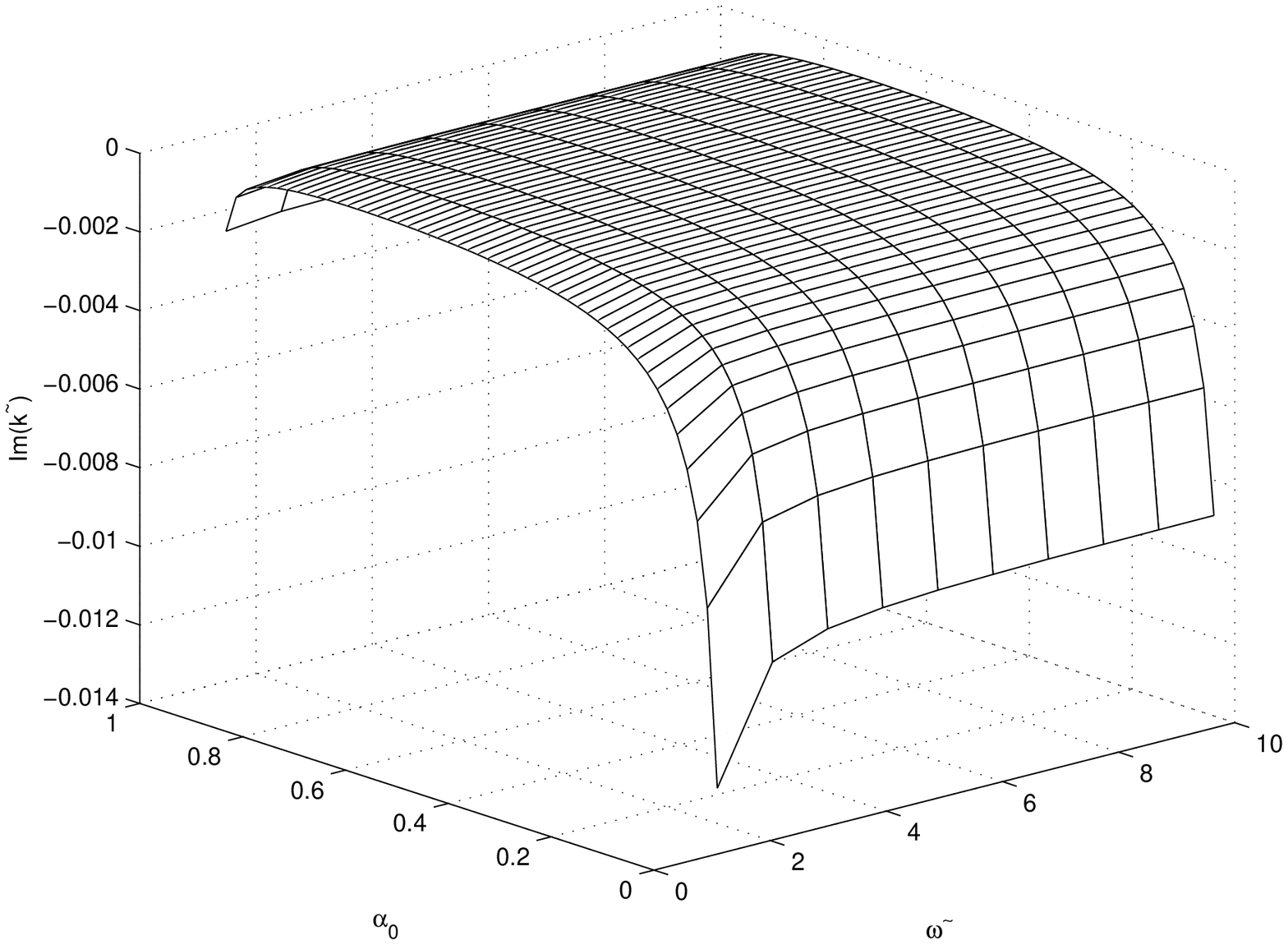}\\
 \includegraphics[scale=.34]{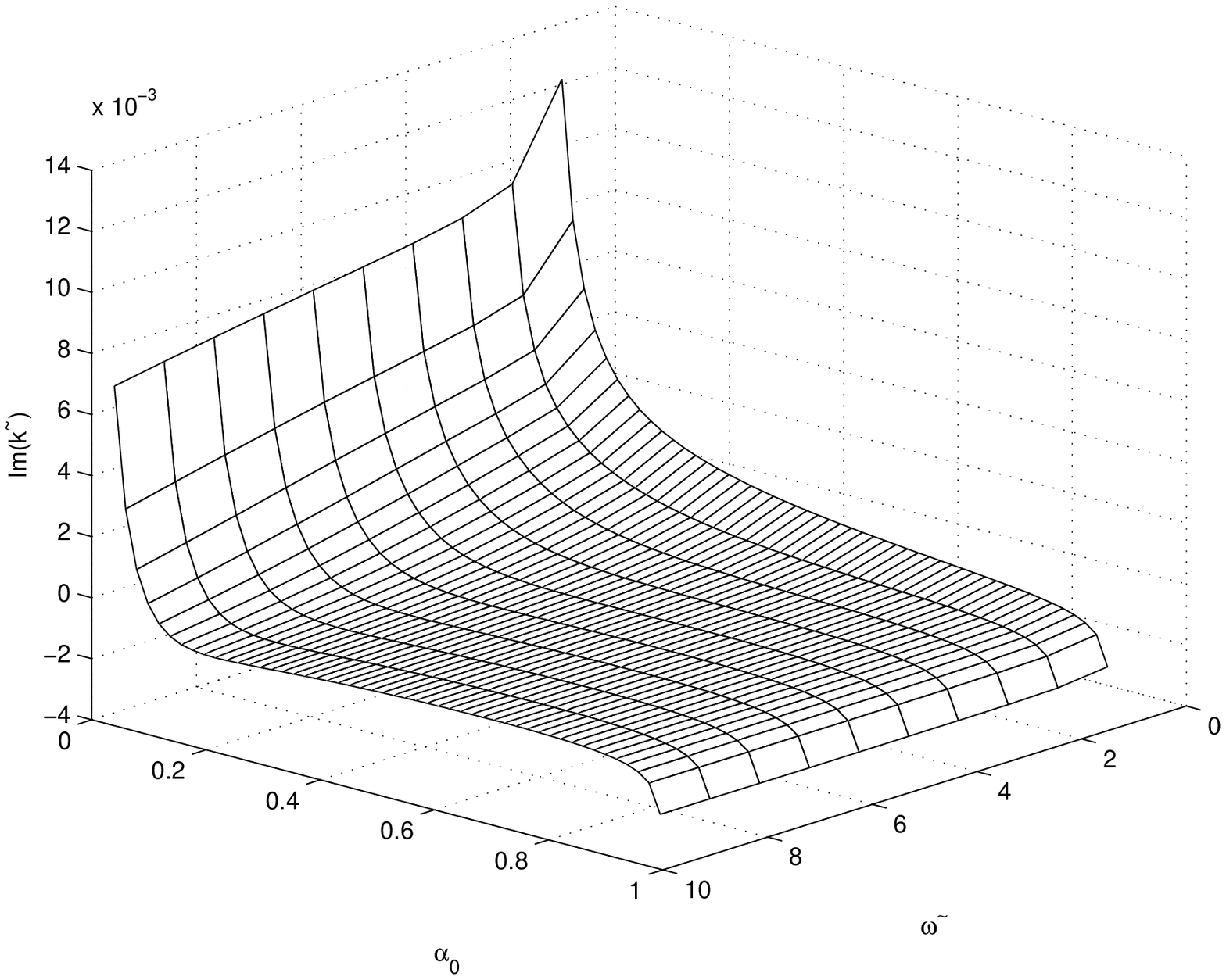}
\end{center}
\caption{\it Top: Real part of the two high frequency growth and both growth and damping modes for the electron-positron plasma. Center: Imaginary part of the growth mode. Bottom: Imaginary part of the mode showing both damping and growth.}
\end{figure}
\centerline{\large{\it{\bf 2. Electron-ion Plasma}}}
Like the electron-positron plasma, the electron-ion plasma admits four high frequency modes. Fig. 7 shows two modes as electron-positron plasma. The mode shown in the center is a growth mode but the bottom mode is growth all the frequency and for  $\alpha_o>0.1$, but shows growth very closed to the black hole horizon. Thus closed to the horizon corresponding to $\alpha _o< 0.1$, it appears that energy is no longer fed into wave mode by the gravitational field but begins to be drained from the waves. The third mode shown in the center of Fig. 8 is damped for all the frequency domain but the fourth mode shown in the bottom of Fig. 8 is damped for most of the frequency domain and for $\alpha_o>0.1$ and shows growth for lower frequencies and for $\alpha _o<0.1$. From these four modes it is clear that in the special relativistic limit they coalesce into only one purely high frequency mode in the special relativistic case investigated by SK\cite{thirty one} for the ultrarelativistic electron-positron plasma. The growth and decay rates obviously depend on frequency, unlike the corresponding Alfv\'en modes.
\begin{figure}[h]\label{fig6}
\begin{center}
 \includegraphics[scale=.34]{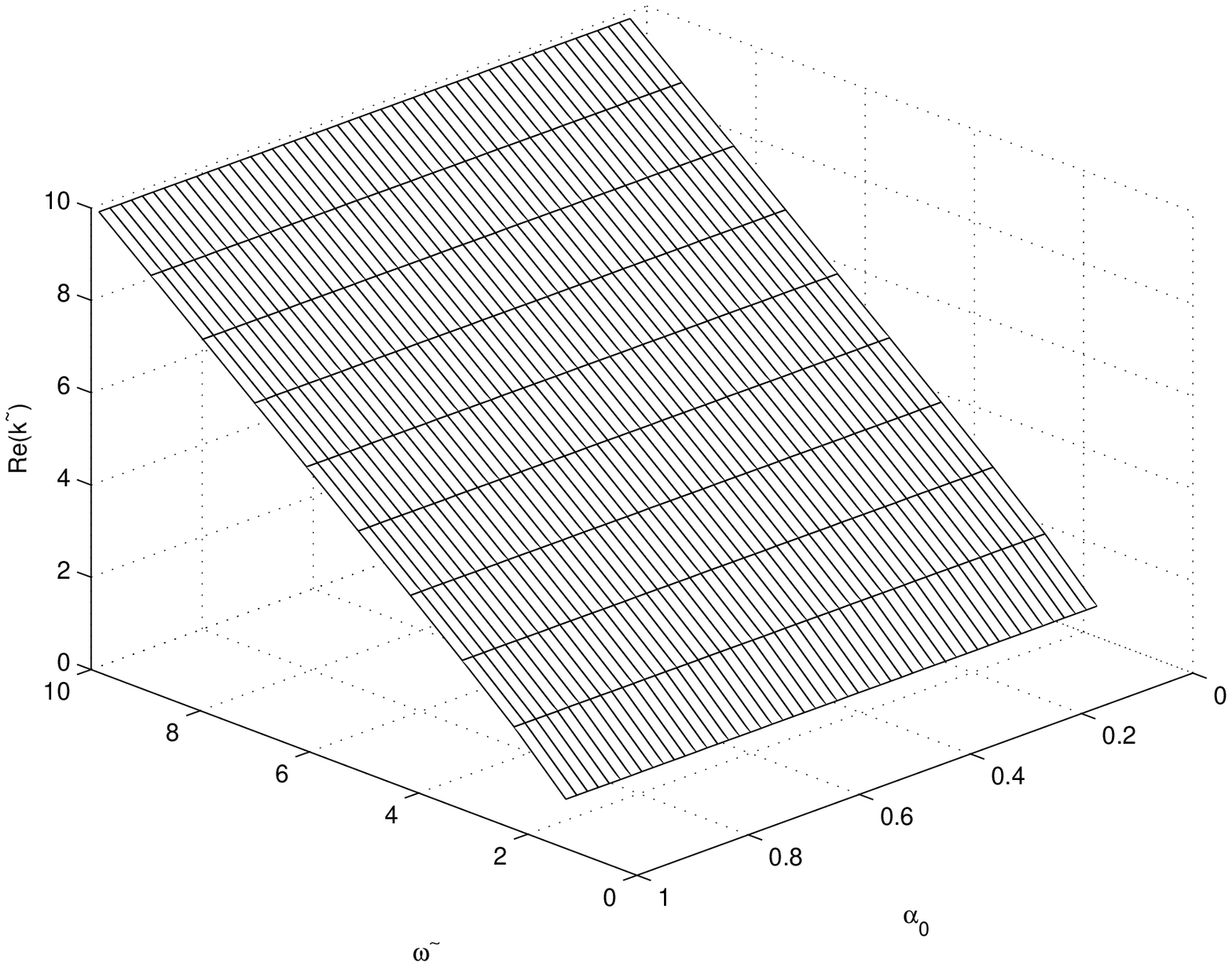}\\
 \includegraphics[scale=.34]{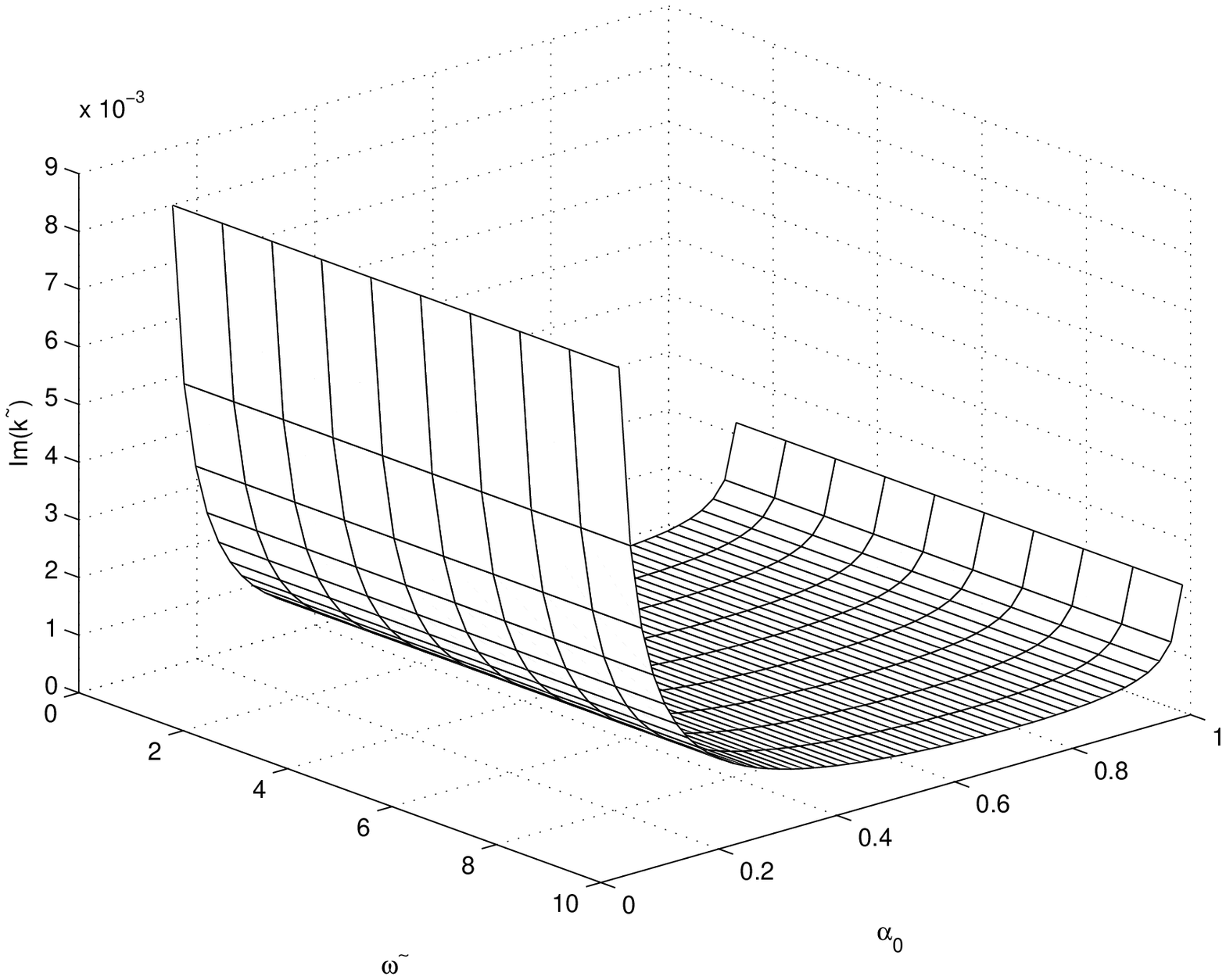}\\
 \includegraphics[scale=.34]{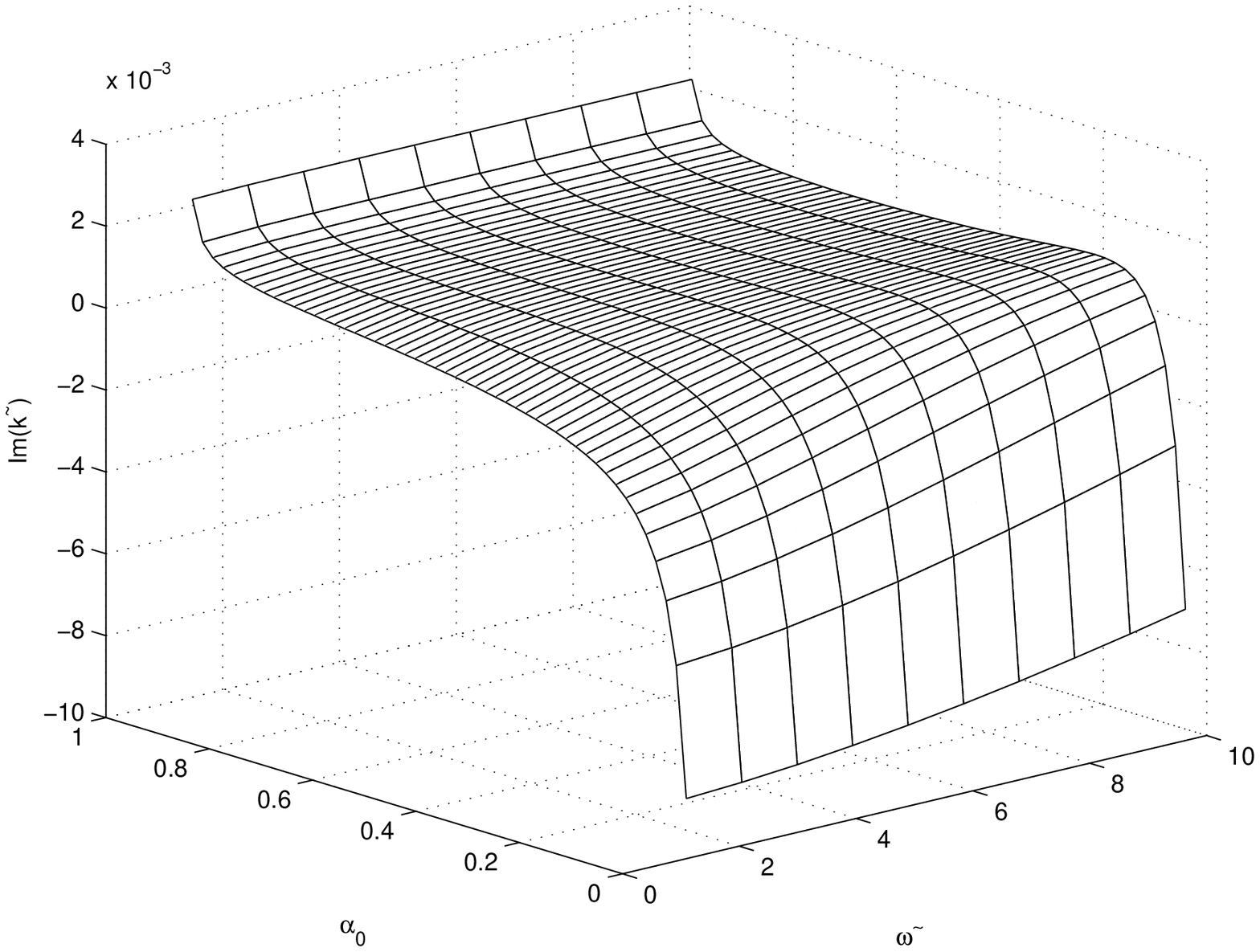}
\end{center}
\caption{\it Top: Real part of the two high frequency damping and both damping and growth modes for the electron-positron plasma. Center: Imaginary part of the damped mode. Bottom: Imaginary part of the damping and growth mode.}
\end{figure}

\section{Concluding Remarks}\label{sec10}

In this paper the linearized two-fluid equations are used to obtain the dispersion relation for transverse waves closed to the event horizon of Schwarzschild Anti-de Sitter black hole. In the limit of zero gravity our results reduce to those in special relativity obtained by Sakai and Kawata \cite{thirty one} where only one purely real mode was found to exist for both the Alfv\'en and high frequency electromagnetic waves for ultra-relativistic electron-positron plasma.
\begin{figure}[h]\label{fig7}
\begin{center}
 \includegraphics[scale=.34]{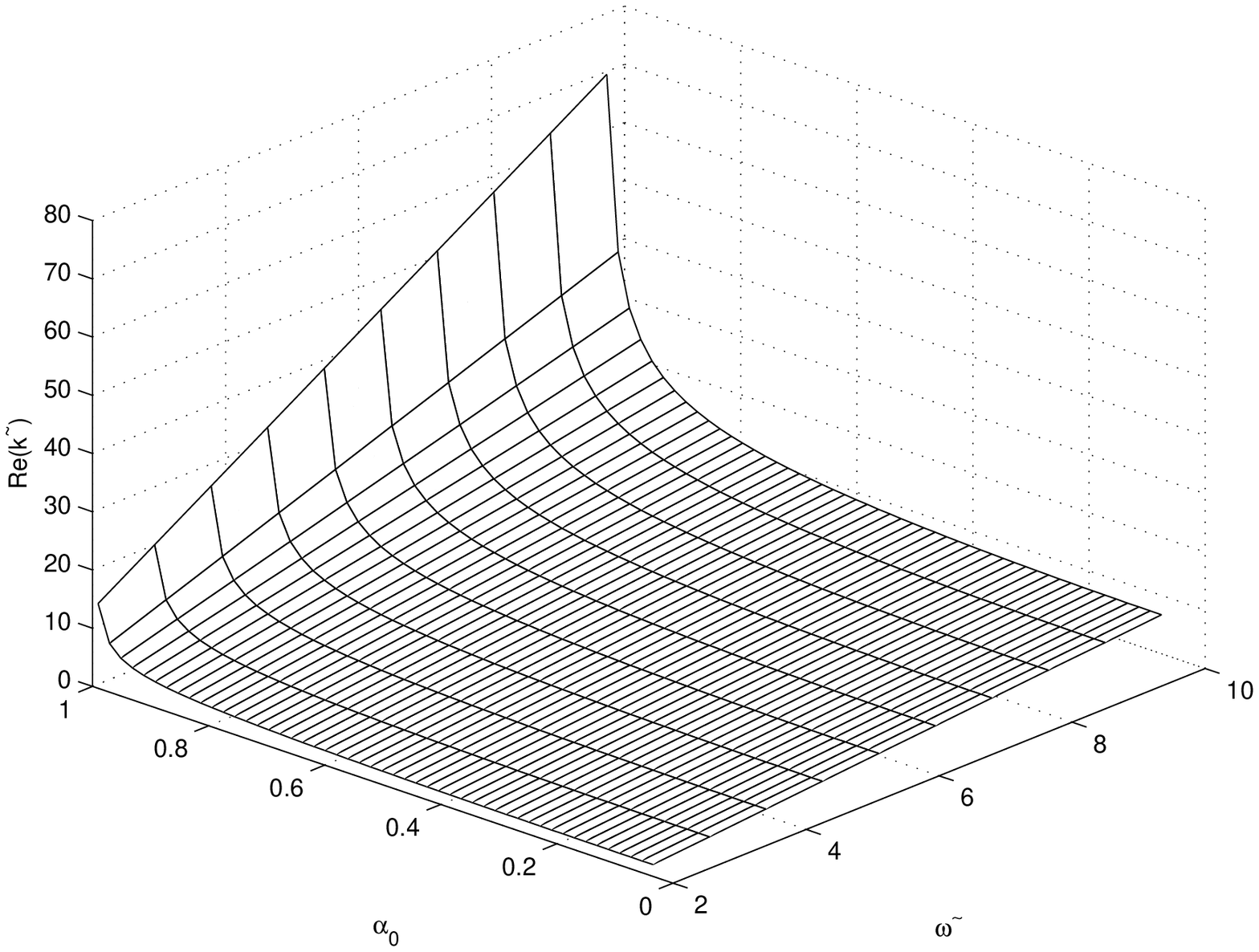}\\
 \includegraphics[scale=.34]{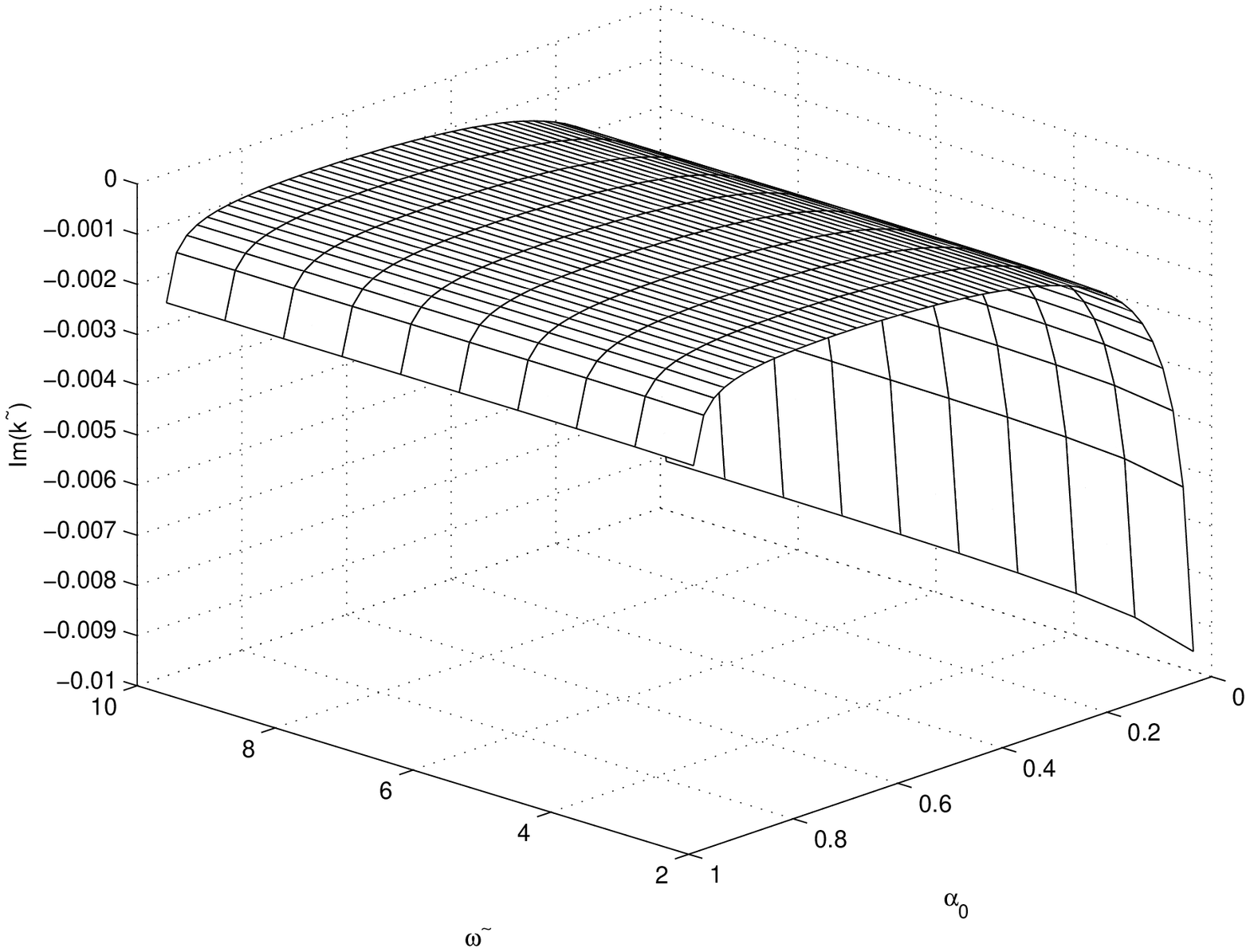}\\
 \includegraphics[scale=.34]{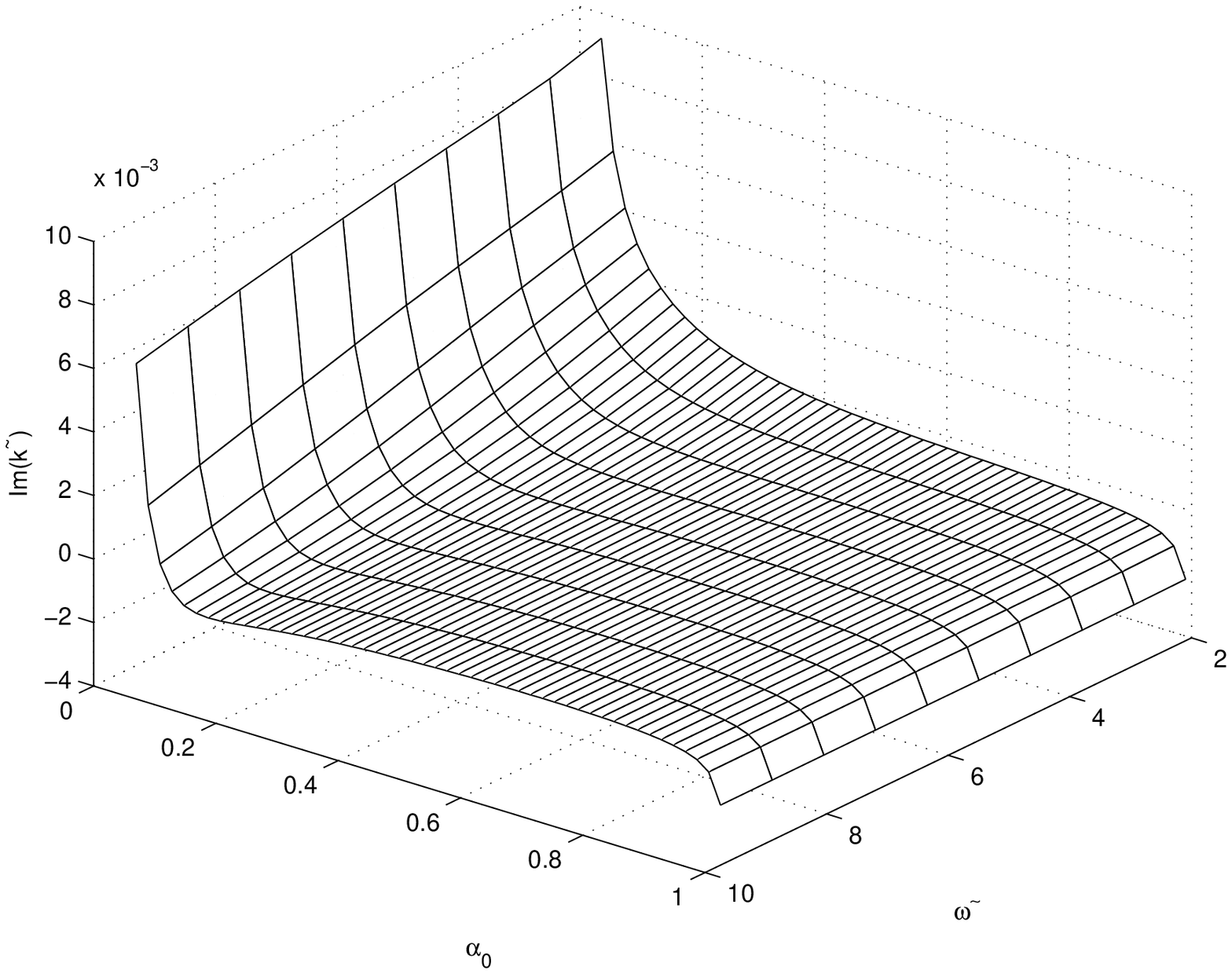}
\end{center}
\caption{\it Top: Real part of the two high frequency growth and both growth and damping modes for the electron-ion plasma. Center: Imaginary part of the growth mode. Bottom: Imaginary part of both growing and damped mode.}
\end{figure}
In general relativistic limit some new modes arise for the Alfv\'en and high frequency electromagnetic waves due to the black hole's gravitational field. For the electron-positron plasma, the damping and growth rates are smaller in comparison with the modes for Schwarzschild black hole by several orders of magnitude, compared with the real components of the wave number. From the various modes for each of the wave types, the damping and growth rates are smaller. A change in the radius of the black hole, however, does not make any significant difference in the result for the field perturbations as it is clear from the high frequency waves shown for each fluid component.
\begin{figure}[h]\label{fig8}
\begin{center}
 \includegraphics[scale=.34]{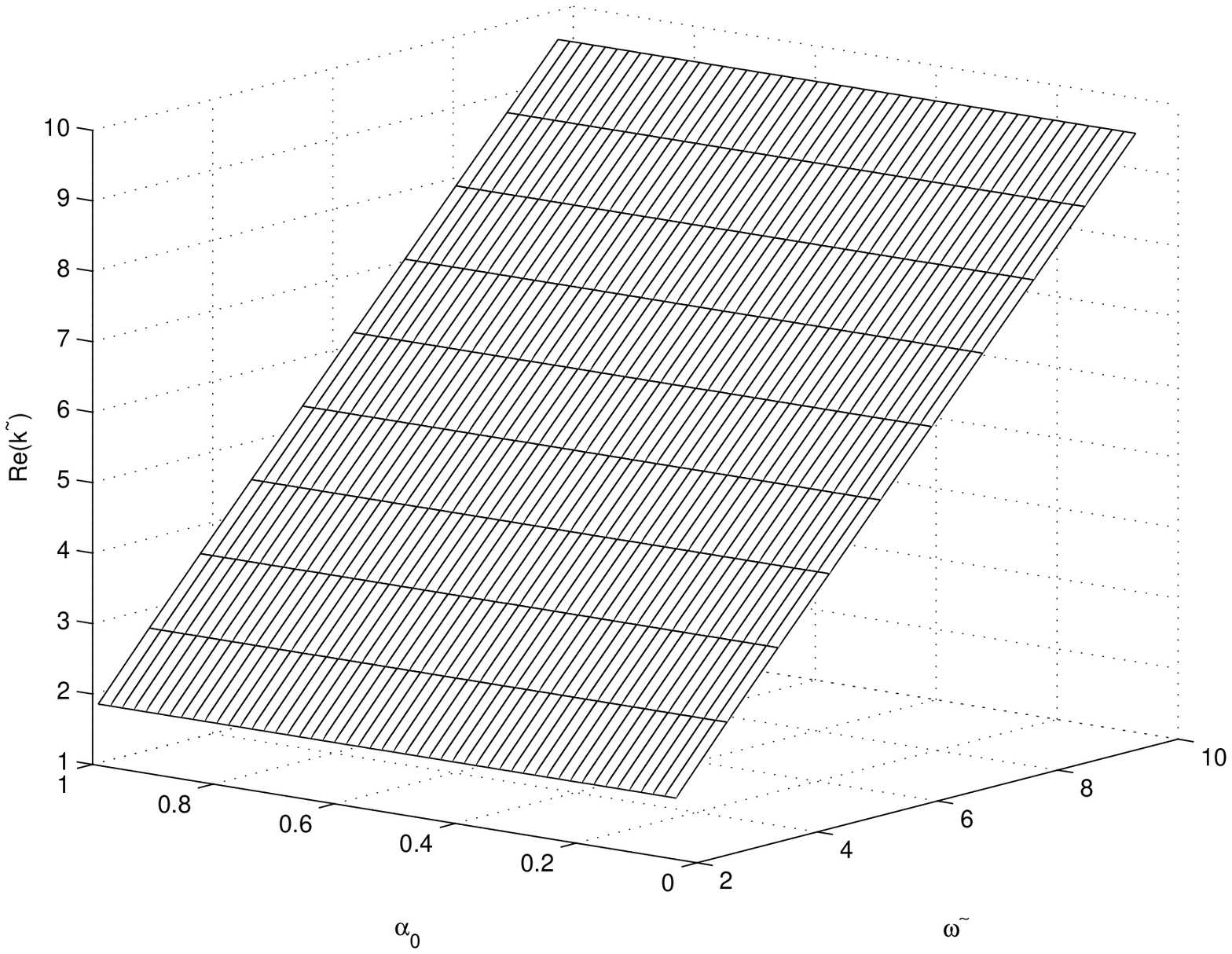}\\
 \includegraphics[scale=.34]{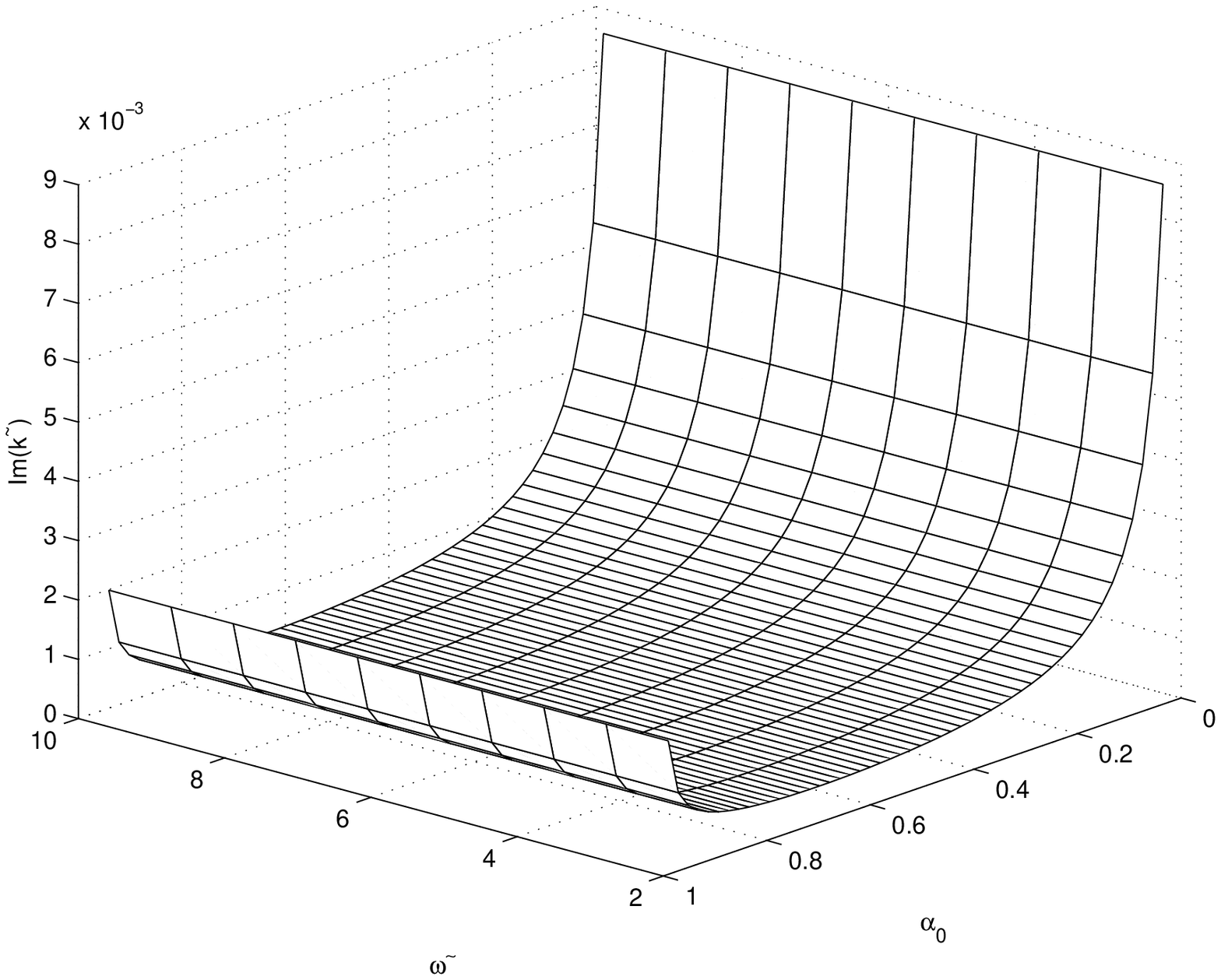}\\
 \includegraphics[scale=.34]{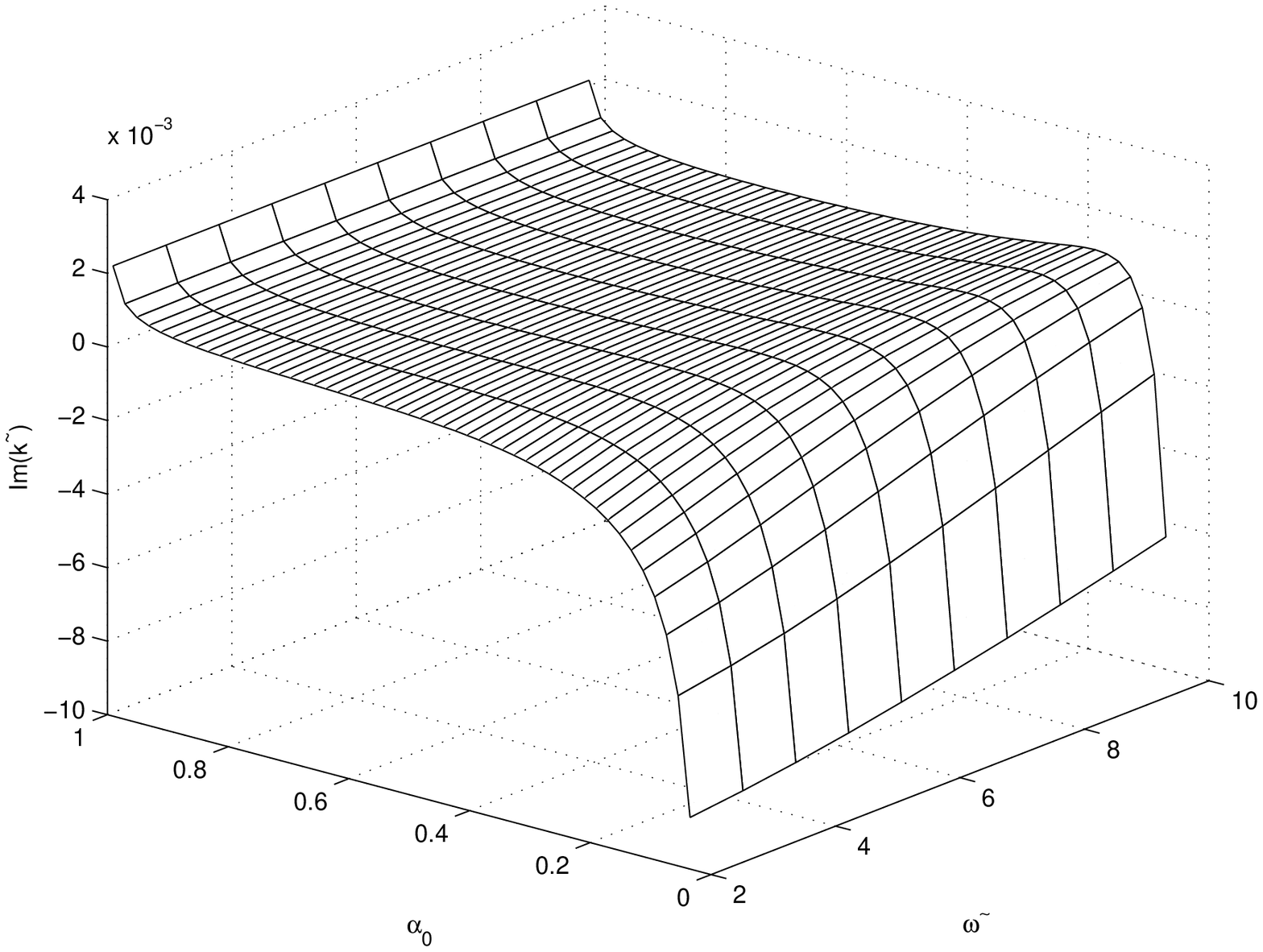}
\end{center}
\caption{\it Top: Real part of the two high frequency damping and both damping and growth modes for the electron-ion plasma. Center: Imaginary part of the damped mode. Bottom: Imaginary part of the damping and growth mode.}
\end{figure}
For the Alfv\'en waves, the damping and growth rates are obviously frequency independent, but are dependent on the radial distance from the horizon as denoted by the mean value of the lapse function $\alpha _0$. This is of course not the case for the high frequency waves. In that case the rate of damping or growth is dependent on both frequency and radial distance from the horizon. So we make conclude that the damping modes demonstrates the energy drained from the waves by the gravitational field and the growth rates indicate that the gravitational field is, in fact, feeding energy into the waves. In the limit $\ell^2\rightarrow\infty$ our study provides the results for the Schwarzschild black hole case \cite{six}.
 
For a multi-component charged plasma in the presence of electromagnetic fields, Marklund et al \cite{thirty two} have analyzed an exact 1+3 covariant dynamical fluid equations in a Friedmann-Robertson-Walker universe and then have considered the matter field of an ion-electron plasma with zero average pressure to an Einstein-de Sitter universe. The cyclotron damping of the gravitational wave produced from resonant interaction with (collisionless) plasma particles have also studied by Servin et al \cite{thirty three} and Kleidis et al \cite{thirty four,thirty five}, while  the nonlinear interactions between gravitational radiation and modified Alfv\'{e}n modes have investigated by Forsberg et al \cite{thirty six} in astrophysical dusty plasmas.

In the context of cosmology, there are also works on fluid dynamics and kinetic gas theory such as gas kinetics in the Friedmann-Lema\^{\i}tre-Robertson-Walker (FLRW) mode  cite{thirty one}, the behavior of matter in the presence of electromagnetic fields with plasma effects \cite{thirty seven,thirty eight,thirty nine,fourty,fourty one,fourty two,fourty three}. Therefore, the general relativistic treatment of plasmas, both in astrophysics as well as in cosmology, seems to be a field open to investigation.

\vspace{0.5cm}
\noindent
{\large\bf Acknowledgement}\\
The authors (MAR) thanks the Abdus Salam International Centre for Theoretical Physics (ICTP), Trieste, Italy, for giving opportunity to utilize its e-journals for research purpose.


\begin{thebibliography}{99}
\bibitem{one}
T. Vachaspati, D. Stojkovic, and L.M. Krauss: \textit{Phys. Rev. D}
\textbf{76}, (2007)024005; [gr-qc/0609024].
\bibitem{two}
H. Ohaninan, R. Ruffin: \textit{Gravitation and Spacetime (New York:Norton)},(1994)\textbf{484}.
\bibitem{three}
D. B. Cling, D. A. Sanders and H. Hong:\textit{ApJ,}
\textbf{486}, (1997)169.
\bibitem{four}
A. F. Heckler: \textit{Phys. Rev. Lett.} \textbf{55}, (1997)480.
\bibitem{five}
A. F. Heckler: \textit{Phys. Rev. Lett.} \textbf{78}, (1997)3430.
\bibitem{six}
Buzzi, V., Hines, K.C., and Treumann, R.A. (1995). {\it Physical Review} D {\bf 51}, 6663.
\bibitem{seven}
Buzzi, V., Hines, K.C., and Treumann, R.A. (1995). {\it Physical Review} D {\bf 51}, 6677.
\bibitem{eight}
Thorne, K.S., and Macdonald, D.A. (1982). {\it Mon. Not. R. Astron. Soc.} {\bf 198}, 339.
\bibitem{nine}
Macdonald, D.A., and Thorne, K.S. (1982). {\it Mon. Not. R. Astron. Soc.} {\bf 198}, 345.
\bibitem{ten}
Price, R.H., and Thorne, K.S. (1986). {\it Physical Review} D {\bf 33}, 915.
\bibitem{eleven}
Thorne, K.S., Price, R.H., and Macdonald, D.A. (1986). {\it Black Holes: The Membrane Paradigm}, Yale University Press, New Haven.
\bibitem{twelve}
Zhang, Xi.-H. (1989). {\it Physical Review} D {\bf 39}, 2933.
\bibitem{thirteen}
Zhang, Xi.-H. (1989). {\it Physical Review} D {\bf 40}, 3858.
\bibitem{fourteen}
Holcomb, K.A., and Tajima, T. (1989). {\it Physical Review} D {\bf 40}, 3809.
\bibitem{fifteen}
Holcomb, K.A. (1990). {\it Astrophysical Journal} {\bf 362}, 381.
\bibitem{sixteen}
Dettman, C.P., Frankel, N.E., and Kowalenko, V. (1993). {\it Physical Review} D {\bf 48}, 5655.
\bibitem{seventeen}
Arnowitt, R., Deser, S., and Misner, C.W. (1962). in {\it Gravitation: An Introduction to Current Research}, edited by Witten, L., (Wiley, New York).
\bibitem{eighteen}
Evans, C.R., Smarr, L.L., and Wilson, J.R. (1986). in {\it Astrophysical Radiation Hydrodynamics}, edited by Norman, M., and Winkler, K.H., (Reidel, Dordrecht).
\bibitem{nineteen}
J. Maldacena, \textit{Adv. Theor. Math. Phys.}
\textbf{2}, (1998)231.
\bibitem{twenty}
E. Witten, \textit{Adv. Theor. Math. Phys.}
\textbf{2}, (1998)253.
\bibitem{twenty one}
S. Kalyana Rama and Sathiapalan,\textit{Mod.Phys. Lett. A}
\textbf{14}, (1999)2635.
\bibitem{twenty two}
E. Witten: \lq \lq Quantum Gravity in de Sitter Space\rq \rq ,
hep-th/0106109.
\bibitem{twenty three}
A. Strominger: \textit{JHEP} \textbf{10}, (2001)034.
\bibitem{twenty four}
A. Strominger: \textit{JHEP} \textbf{11}, (2001)049.
\bibitem{twenty five}
C.M. Hull: \textit{JHEP} \textbf{07}, (1998)021.
\bibitem{twenty six}
C.M. Hull, and R.R. Khuri: \textit{Nucl. Phys. B} \textbf{575},
(2000)231.
\bibitem{twenty seven}
P.O. Mazur, and E. Mottola: \textit{Phys. Rev. D} \textbf{64},
(2001)104022.
\bibitem{twenty eight}
M. Colpi, L. Maraschi, and A. Treves: \textit{Astrophys. J.}
\textbf{280}, (1984)319.
\bibitem{twenty nine}
E. Harris: \textit{Phys. Rev.} \textbf{108}, (1957)1357.
\newpage
\bibitem{thirty}
F. J\"uttner: \textit{Ann. Phys.} (Leipzig) \textbf{34}, (1911)856.
\bibitem{thirty one}
J. Sakai, and T, Kawata,(1980). {\it Journal of Physical Society in Japan} {\bf 49}, 747.
\bibitem{thirty two}
M. Marklund, P.K.S. Dunsby, G. Betschart, M. Servin, and C.G. Tsagas: \textit{Class. Quantum Grav.} \textbf{20}, (2003)1823.
\bibitem{thirty three}
M. Servin, and G. Brodin: \textit{Phys. Rev. D} \textbf{64}, (2003)024013.
\bibitem{thirty four}
K. Kleidis, H. Varvoglis, D. Papadopoulos, and F.P. Esposito: \textit{Astron. Astrophys.} \textbf{294}, (1995)313.
\bibitem{thirty five}
K. Kleidis, H. Varvoglis, and D. Papadopoulos: \textit{Class. Quantum Grav.} \textbf{13}, (1996)2547.
\bibitem{thirty six}
M. Forsberg, G. Brodin, M. Marklund, P.K. Shukla, and J. Moortgat: \textit{Phys. Rev. D} \textbf{74}, (2006)064014.
\bibitem{thirty seven}
J. Bernstein: Kinetic Theory in the Expanding Universe (Cambridge: Cambridge University Press) 1988.
\bibitem{thirty eight}
D. Papadopoulos, and F.P. Esposito: \textit{Astrophys. J.} \textbf{257}, (1982)10.
\bibitem{thirty three}
K. Subramanian, and J.D. Barrow: \textit{Phys. Rev. D} \textbf{58}, (1998)083502.
\bibitem{thirty nine}
C.G. Tsagas, and J.D. Barrow: \textit{Class. Quantum Grav.} \textbf{14},  (1997)2539.
\bibitem{fourty}
C.G. Tsagas, and J.D. Barrow: \textit{Class. Quantum Grav.} \textbf{15},  (1998)3523.
\bibitem{fourty one}
C.G. Tsagas, and R. Maartens: \textit{Phys. Rev. D} \textbf{61}, (2000)083519.
\bibitem{fourty two}
C.G. Tsagas, and R. Maartens: \textit{Class. Quantum Grav.} \textbf{17}, (2000)2215.
\bibitem{fourty three}
M. Marklund, P.K.S. Dunsby, and G. Brodin: \textit{Phys. Rev. D} \textbf{62}, (2000)101501.

\end{thebibliography}
\end{document}